\documentstyle[prd,aps,preprint,amsfonts]{revtex}
\begin{document}
\tighten
%%%%%%%%%%%%%%%%%%%%%%%%%%%%%%%%%%%%%%%%%%%%%%%%%%%%%%%%%%%%%
\typeout{--------------------------------------------------}
\typeout{If you have trouble printing this document, please}
\typeout{obtain the compiled dvi or ps-file by anonymous ftp}
\typeout{to ftp://tpri6c.gsi.de/pub/phenning/h95spec/ or at}
\typeout{http://www.gsi.de/<tilde>phenning/tft.html}
\typeout{--------------------------------------------------}
%%%%%%%%%%%%%%%%%%%%%%%%%%%%%%%%%%%%%%%%%%%%%%%%%%%%%%%%%%%%%%%%%%%%%%
\title{Fermion damping rate in a hot medium
       \thanks{Work supported by GSI.
GSI-Preprint 94-56 revised,
 subm. to Phys.Rev. D (hep-ph 9409280)}}
\author{P.A.Henning, R.Sollacher}
\address{Theoretical Physics,
        Gesellschaft f\"ur Schwerionenforschung GSI\\
        P.O.Box 110552, D-64220 Darmstadt, Germany\\
        P.Henning@gsi.de}
\author{H.Weigert}
\address{Theoretical Physics Institute\\
         116 Church St. SE, Minneapolis MN 55455\\
         hweigert@physics.spa.umn.edu}
\maketitle
\begin{abstract}
In principle every excitation acquires a finite lifetime in a hot system.
This nonzero spectral width is calculated self-consistently
for massive fermions coupled to massless scalar, vector and
pseudoscalar bosons. It is shown that the self-consistent summation
of the corresponding Fock diagram for fermions eliminates all
infrared divergences although the bosons are not screened at all.
Our solutions for the fermion
damping rate are analytical in the coupling constant,
but not analytical in the temperature parameter around $T=0$.
\end{abstract}
%%%%%%%%%%%%%%%%%%%%%%%%%%%%%%%%%%%%%%%%%%%%%%%%%%%%%%%%%%%%%%%%%%%%%%
\section{Introduction}
Recently, the problem of the {\em damping rate\/} of a fermion moving
through a hot system, i.e., the width of its spectral function,
has attracted much attention \cite{LS,P93,BNN93,PPS93,BPS94,BR94,W94,BK94,T94}.
This damping rate determines the radiation spectrum \cite{hq95gam}
as well as the non-equilibrium transport
properties of a relativistic plasma \cite{h95neq}, hence is
an interesting quantity for a wide range of physical systems:
Relativistic heavy-ion collisions, fusion reactors
and the early universe are some examples.

The damping rate of fermions interacting with {\em massless\/} bosons
is an especially interesting problem in this framework.
{}From the phenomenological point of view this importance is due to the fact
that all fundamental forces of nature involve massless gauge bosons.
{}From the theoretical standpoint a similarly big interest
is due to the calculational difficulty associated with this
particular problem.

This difficulty is an infrared divergence, i.e.,
if interacting with strictly massless bosons a fermion may emit
an infinite number of such bosons with infinitely small energy.
Such a concept is of course meaningless in the real world, and
consequently the solution of the problem has to connect the
abstract world of quantum physics with reality. We may indeed
consider the solution as almost crucial for our understanding of nature
by means of theoretical modeling.

Such an infrared divergence occurs already at zero temperature
in QED as well as in QCD. The connection to reality we mentioned,
and which is conventionally used to remove this unphysical feature,
is the finite energy resolution of every physical measuring apparatus.

At nonzero temperature however, the question of the gauge-independent,
causality preserving and therefore {\em physical\/} removal
of the infrared divergence is far from settled. With the present paper
we wish to contribute a new answer to this question, based
on a perturbative expansion in terms of unstable asymptotic states.

Our contribution is complementary to the most common method
to remove infrared divergences at nonzero temperature, i.e.,
to the method of hard thermal loops \cite{BP90}. In this framework
a partial resummation of diagrams contributes a small ``thermal''
mass to particles, whereas their thermal scattering is usually
neglected. However, this thermal scattering, or Brownian motion,
is the fundamental difference between a system at nonzero temperature
and a system at $T=0$, since it leads to a finite lifetime of
{\em any\/} excitation. Indeed,
one of the few mathematically rigorous theorems of finite temperature
quantum theory, due to Narnhofer and Thirring, has established this
once and for all \cite{NRT83}: ''Particles'' with infinite lifetime
can exist in a hot system only if they do not interact.
The use of particle-like excitations in a perturbative scheme,
which are asymptotically stable
and only have a temperature dependent ''mass'' is therefore unjustified
{\em a priori\/} and can only be obtained as a limiting case
at the very end of all calculations.

The deeper mathematical reason is that the symmetry group of
space and time in thermal states is not the Poincar\'e group as in the vacuum
state, but rather a product of SO(3) and the four-dimensional translation
group \cite{BS75}. Its irreducible representations can be interpreted
as having a continuous mass spectrum, i.e., the concept of a
mass-shell (and thereby of stable asymptotic states)
is not well-defined at nonzero temperature.

The solution of this problem has been pointed out by Landsman \cite{L88},
using earlier work of Licht \cite{L65} and Wightman \cite{W67}
and leading to a perturbative expansion in terms of
{\em generalized free fields\/} without a mass-shell. To account for
the proper temporal boundary conditions in a thermal system,
these generalized free fields have to be embedded
in a description with doubled Hilbert space, i.e.,
the single-particle propagators are $2\times 2$ matrices \cite{LW87}.
Two flavors exist for such a formalism, the Schwinger-Keldysh (or
closed time-path) method \cite{SKF} and the method called
thermo field dynamics (TFD) \cite{Ubook}.

Within the latter method this merger of different aspects of finite
temperature field theory has been discussed in a recent overview
\cite{h94rep}, we will follow this description closely with our notation.
Although in general we prefer TFD over the closed time-path method due to its
technical simplicity, the present paper is formulated in
a way that is independent of the choice of the formalism.

Within this framework, we now consider the problem of the fermion damping
rate. The paper is organized as follows: First, we make some general
remarks on propagators in terms of their spectral function. These
remarks are then used to obtain well-defined approximations to the
spectral functions of bosons and fermions in a hot medium.

In section III we briefly discuss the problem of calculating
one-loop diagrams with effective propagators, and how they may be used
for a self-consistent determination of the spectral function.
In the following we drop some of the previous results,
concentrating on the fermion spectral width only:
Sections \ref{GamVac} and \ref{GamTem} are devoted to analytical
approximations for the Fock self energy function.

In section \ref{scfa} we assemble these results, and
consequently obtain approximate solutions for the self-consistent
Fock approximation, i.e., for the full fermion propagator
calculated with a one-loop self energy function at finite
temperature. Finally these results, which are different from those obtained
by other methods, are used to draw some conclusions.
%%%%%%%%%%%%%%%%%%%%%%%%%%%%%%%%%%%%%%%%%%%%%%%%%%%%%%%%%%%%%%%
\section{Spectral functions}
As appropriate for a description of dynamical phenomena in a
thermal system, we employ a formalism with $2\times 2$ matrix valued
Green functions. In thermo field dynamics (TFD) this ``doubling''
arises naturally because TFD contains two different (anti-) commuting
representations of the canonical (anti-) commutation relations.

In the Schwinger-Keldysh method the doubling is introduced
because a full description of statistical systems can be achieved
only when using causal and  anti-causal Green functions as well as
Wigner functions. Consequently, the full propagator matrix is
\begin{equation} \label{cttp}
S^{(ab)}(x,x^\prime)  = -{\mathrm i}\left( {
   \array{lr}
    \left\langle{\mathrm T}\left[\psi_x \overline{\psi}_{x^\prime}
    \right]\right\rangle&
   -\left\langle\overline{\psi}_{x^\prime}\psi_x \right\rangle\\[2mm]
    \left\langle\psi_x\overline{\psi}_{x^\prime}\right\rangle&
    \left\langle\widetilde{{\mathrm T}}\left[\psi_x\overline{\psi}_{x^\prime}
    \right]\right\rangle
   \endarray} \right)
\;,\end{equation}
where $\left\langle\cdot\right\rangle$
denotes the statistical average, ${\mathrm T}[\cdot]$
the time ordered and $\widetilde{{\mathrm T}}[\cdot]$ the
``anti-time ordered'' product. The same matrix valued propagator
is obtained in TFD (see \cite{h94rep,hu92} for detailed
discussions of this equivalence).

By construction the propagator matrix obeys the linear relation
$S^{11}+S^{22}-S^{12}-S^{21} = 0$. Furthermore, in an equilibrium state
it has to fulfill the Kubo-Martin-Schwinger condition
\cite{KMS}, i.e., an anti-periodic boundary condition in the imaginary time
direction. We will henceforth formulate all our considerations in momentum
space, the KMS condition then reads \cite{LW87,h94rep,hu92}
\begin{equation}\label{kmf}
\left(1 - n_F(p_0)\right)S^{12}(p_0,\bbox{ p}) +
 n_F(p_0) S^{21}(p_0,\bbox{ p}) = 0
\;.\end{equation}
$S^{12}$ and $S^{21}$ are the Green functions without time ordering
(Wigner functions) and $n_F(E)$ is the fermion equilibrium
distribution function at a given temperature, the Fermi-Dirac function
\begin{equation}\label{nfdf}
n_F(E) = \frac{1}{\displaystyle\mbox{e}^{ \beta E}+1}
\;.\end{equation}
An almost identical KMS relation holds for bosons, but here the
fields are periodic in the imaginary time direction.
Consequently,
occupation number factors are given as Bose-Einstein function
\begin{equation}\label{nbdf}
n_B(E) = \frac{1}{\displaystyle\mbox{e}^{\beta E}-1}
\;.\end{equation}
Henceforth for simplicity
we consider only the case of zero chemical potential.

In interacting systems, the KMS boundary conditions lead to a
{\em causal\/} propagator which has a cut along the real energy axis.
Its analytical structure therefore is not easily understood,
and especially when combining several of such propagators in a
perturbative scheme one has to implement more or less complicated
cutting rules to understand pieces of diagrams {\em physically\/}
\cite{KS85c,KV95}.

Although this may be a matter of personal taste, we therefore
deem it the safest method to use only retarded and advanced
propagators, and to condense the matrix structure of eq.
(\ref{cttp}) into the vertices where such propagators join.
This amounts to a {\em diagonalization scheme\/} for the matrix
(\ref{cttp}), which has been described in several publications
\cite{h94rep,hu92}.
%%%%%%%%%%%%%%%%%%%%%%%%%%%%%%%%%%%%%%%%%%%%%%%%%%%%%%%%%%%%%%%%%%%%%%
\subsection{Retarded and advanced propagators}
The retarded and advanced propagators
are, by definition, analytical functions of the
energy parameter in the upper or lower complex energy half plane.
Analytical functions obey the Kramers-Kronig relation,
and this implies that the retarded propagator is known completely
if only its imaginary part (or spectral function) ${\cal A}_F$ is known
along the real axis. Hence, for arbitrary complex $E$
\begin{equation}\label{rapf}
S^{R,A}(E,\bbox{ p})  =
  \int\limits_{-\infty}^{\infty}\!\!dE^\prime\;
  {\cal A}_F(E^\prime,\bbox{ p})\;
   \frac{1}{E-E^\prime\pm{\mathrm i}\epsilon}
\;.\end{equation}
Trivially, the spectral function is recovered for real $E$,
\begin{equation}\label{gsp}
{\cal A}_F(E,\bbox{ p}) = \mp\frac{1}{\pi}\,\mbox{Im}(S^{R,A}(E,
\bbox{ p}))
  = \frac{1}{2\pi{\mathrm i}}\left(S^A(E,\bbox{ p}) - S^R(E,
\bbox{ p})\right)
\;.\end{equation}
We consider the diagrammatic rules to combine these
propagators in the calculation of physical quantities to be well
established. For completeness we summarize the diagonalization
of eq. (\ref{cttp}) in terms of the spectral function
${\cal A}_F(E,\bbox{ p})$ as \cite{h94rep,hu92}
\begin{eqnarray}\nonumber
&&S^{(ab)}(p_0,\bbox{ p}) =
   \int\limits_{-\infty}^\infty\!\!dE\,
        {\cal A}_F(E,\bbox{ p})\;\times \\ \label{fsk1}
&&\tau_3\, ({\cal B}(n_F(E)))^{-1}\;
   \left(\!{\array{ll}
         {\displaystyle \frac{1}{p_0-E+{\mathrm i}\epsilon}} & \\
    &    {\displaystyle \frac{1}{p_0-E-{\mathrm i}\epsilon}}
\endarray}\right)\;
 {\cal B}(n_F(E))
\;.\end{eqnarray}
$\tau_3$ is the diagonal Pauli matrix,
and the transformation matrix ${\cal B}$ is
\begin{equation}\label{fex}
{\cal B}(n_F(E)) =   \left(\array{ll}(1-n_F(E)) & -n_F(E)\\
                1 & 1\endarray\right)
\;.\end{equation}
It is a matter of a few lines to show that this matrix-valued
propagator obeys relation (\ref{kmf}) for each spectral function.

A similar relation holds for the boson case,
see \cite{h94rep,hu92} for details of the corresponding ${\cal B}$.
For both cases, the transformation matrices ${\cal B}$ have a special meaning
in the TFD formalism, where they play the role of a thermal
Bogoliubov transformation.
However, the explicit form (\ref{rapf}) of the propagator is
the same in the Schwinger-Keldysh method.
%%%%%%%%%%%%%%%%%%%%%%%%%%%%%%%%%%%%%%%%%%%%%%%%%%%%%%%%%%%%%%%%%%%%%%
\subsection{Normalization and Locality}
The spectral function has two features which are intimately related to
fundamental requirements of quantum field theory.
We will comment on these only for the fermionic case, and refer
to \cite{h95comm} for the case of bosons.
Firstly, the quantization rules for fields
(which we will use in the free as well as in the interacting case),
\begin{equation} \label{ccr}
\left\{\psi(t,\bbox{x}),\psi^\dagger(t,\bbox{y})\right\} =
  \delta^3(\bbox{x}-\bbox{y})
\;\end{equation}
require that the spectral function is {\em normalized\/}
\begin{equation}\label{norm}
\int\limits_0^\infty\!dE\,\mbox{Tr}\left[ \gamma^0\,{\cal A}_F(E,
\bbox{ p})\right]
 = 2
\;.\end{equation}
The second important feature of the spectral function is that
its four dimensional Fourier transform into coordinate space must vanish
for space-like arguments. This is equivalent to the Wightman axiom of
{\em locality\/}, i.e., field operators must
(anti-)commute for space-like separations in Minkowski space \cite{GJ81}:
\begin{eqnarray}\nonumber
C_F(x,y) &=&\vphantom{\int\limits_0^0}\left\langle\left\{\psi(x),
  \psi^\dagger(y)\right\}\right\rangle\\
\nonumber
         &=& \int\!\!\frac{dE\,d^3\bbox{p}}{(2\pi)^3}
  \;{\mathrm e}^{\displaystyle -{\mathrm i}(E(x_0-y_0)-
  \bbox{p}(\bbox{x}-\bbox{y}))}\,{\cal A}_F(E,\bbox{p})\\
&=& 0\;\;\; \mbox{if $x-y$ space-like}
\;.\end{eqnarray}
In an interacting many-body system, we may very well expect
non-locality in a causal sense: Wiggling the system at one side will
certainly influence the other side after some time. The
locality axiom ensures that this influence does not occur over
space-like separations, i.e., faster than a physical signal
can propagate. Thus, to distinguish between the {\em causal\/} non-locality
and the violation of the locality axiom, we will henceforth
denote the latter a violation of {\em causality\/}.

It is trivial to establish that normalization and locality axiom
are satisfied for free fermions of mass $M$.
Defining their on-shell energy as
$
\omega_p = \sqrt{\bbox{ p}^2 + M^2}
$,
the free spectral function is obtained as
\begin{equation}\label{ff}
{\cal A}^0_F(E,\bbox{ p}) =
        \left( E\gamma^0 + \bbox{ p}\bbox{ \gamma} + M\right)\,
        \mbox{sign}(E)\,\delta(E^2-\omega^2_p)
\;.\end{equation}
Without loss of generality we set $y=0$ in the
free fermionic anti-commutator function, and obtain
\begin{eqnarray}\nonumber
C_F^0(x,0) &=& \left(
{\mathrm i}\partial_\mu\gamma^\mu + M\right) Z(x_0,\bbox{x})
\;\;\;\mbox{where}\\
\label{c0def}
Z(x_0,\bbox{x})& =& \frac{ - {\mathrm i}}{4\pi}\;
  \left( \delta(x_0^2-\bbox{x}^2) - \Theta(x_0^2-\bbox{x}^2)
  \frac{M}{2 \sqrt{ x_0^2-\bbox{x}^2 }}\,
  J_1(M \sqrt{ x_0^2-\bbox{x}^2}) \right)
\;,\end{eqnarray}
where $J_1$ is a Bessel function of the first kind.
Clearly this is zero for space-like arguments, i.e., for
$|\bbox{x}| > |x_0|$. However, from this prescription follows
that for a general spectral function the locality axiom is not
automatically guaranteed, a careful check is necessary
in any application involving spectral functions.
%%%%%%%%%%%%%%%%%%%%%%%%%%%%%%%%%%%%%%%%%%%%%%%%%%%%%%%%%%%%%%%%%%%%%%
\subsection{Ghost poles}
For an interacting system, the full fermion propagator
and thus also the spectral function is defined in terms of a self
energy function $\Sigma$:
\begin{equation}\label{sol3}\nonumber
{\cal A}_F(p_0,\bbox{ p})=\mp \frac{1}{\pi}\,\mbox{Im}\left[
  \left(p_\mu\gamma^\mu - M - \Sigma^{R,A}(p_0,\bbox{ p})\right)^{-1}
\right]
\;.\end{equation}
Note, that also the self energy function is Dirac matrix valued, i.e.,
this equation contains the inversion of a $4\times 4$ matrix.

However, the real part of the full propagator is {\em not} given by the
real part of the quantity in square brackets of (\ref{sol3}).
The reason is, that according to
Weinberg's theorem \cite{W60} the self energy function in a relativistic
theory behaves as
\begin{equation}\label{uvd}
   \Sigma(p_0,\bbox{ p})\propto p_0\,\left(\log\left(\frac{p_0}{M}
  \right)\right)^n
\;\;\;\mbox{if}\;\;\;p_0\rightarrow\infty
\;,\end{equation}
i.e., it is more than linearly divergent. This implies, in general, the
presence of unwanted poles. For fermions, these Landau ghost poles
appear at four points in the complex energy plane \cite[p.636]{IZ80}.

Since their appearance contradicts the definition of the retarded/advanced
propagator as being free of poles at least in one half plane, they must
be considered unphysical. For sufficiently large coupling, they
have a big influence on the propagator along the real axis, while for
small coupling they are far away from the physical region.
Hence, their consistent
removal is mandatory for strongly interacting systems, and the only way
to do so is by calculating the real part of the propagator by a dispersion
integral according to (\ref{rapf}).

It was pointed out in ref. \cite{h92fock}, that this indeed conforms to
a {\em non-perturbative\/} correction to the propagator,
in the sense that the correction term is not an
analytical function of the coupling constant. In general, it will be
of order $\exp(-1/g^2)$ instead. Hence, eqn. (\ref{rapf})
is the only safe method to calculate a relativistic propagator, and
the input needed for this calculation is a physically meaningful spectral
function which obeys all the rules discussed here.

To demonstrate briefly {\em how\/} the removal of unphysical
poles is achieved through a dispersion integral,
we discuss a bosonic toy model spectral function
of Lorentzian type which has a peak along the real axis
at some value $\omega_p$
\begin{equation}\label{toy}
{\cal A}_0(E,\bbox{ p}) = \frac{1}{\pi}\,\frac{\gamma_p}{
   \left(E-\omega_p\right)^2+\gamma_p^2}
\;.\end{equation}
By definition, this spectral function may be used only for real
energy arguments, and does not take on negative values. If this were violated,
the probabilistic interpretation of the quantum (field) theory
behind this formulation would be invalid.

The above function has a trivial expansion into simple poles in the complex
plane,
\begin{equation}
{\cal A}_0(E,\bbox{ p}) =
\frac{1}{2\pi {\mathrm i}}\left(
\frac{1}{E-\omega_p- {\mathrm i}\gamma_p}-\frac{1}{E-\omega_p+
{\mathrm i}\gamma_p}\right)
\;.\end{equation}
We therefore calculate the real part of the corresponding propagator
through
\begin{equation}\label{disp}
G^R_0(p_0,\bbox{ p}) = \int\limits_{-\infty}^\infty\!\!dE\,
{\cal A}_0(E,\bbox{ p})\,\frac{1}{p_0-E+{\mathrm i}\epsilon}
\;.\end{equation}
The integrand is expanded into simple poles,
\begin{eqnarray}\nonumber
\frac{1}{E-\omega_p\pm{\mathrm i}\gamma_p}\,\frac{1}{p_0-E+{\mathrm i}
\epsilon}
=&&\\
\frac{1}{p_0-\left(\omega_p\mp{\mathrm i}\gamma_p\right)}&&\left(
 \frac{1}{E-\omega_p\pm{\mathrm i}\gamma_p} + \frac{1}{p_0-E+{\mathrm i}
\epsilon}\right)
\;.\end{eqnarray}
The $E$-integration is convergent, and closing the integral in the upper
complex energy half plane then picks up the residues of the corresponding
poles. First consider the case $\mbox{Im}(p_0)\ge 0$, where the equal sign
is permitted because of the ${\mathrm i}\epsilon$ terms in the denominator of
the dispersion integral. Clearly, the integrand then has three poles in the
upper complex half plane, but two of the residues cancel.

In the second case $\mbox{Im}(p_0)<0$ the integrand has only one
pole in the upper complex half plane. Assembled together the integral
gives
\begin{equation}
G^R_0(p_0,\bbox{ p}) =
\left\{ {\array{lll} \displaystyle
\frac{1}{p_0-\left(E_p-{\mathrm i}\gamma_p\right)} & \mbox{if} &
\mbox{Im}(p_0)\ge 0\\[3mm]
\displaystyle
\frac{1}{p_0-\left(E_p+{\mathrm i}\gamma_p\right)} & \mbox{if} &
\mbox{Im}(p_0)< 0
\endarray}\right.
\;,\end{equation}
which is free of poles on both sides of the real axis. A similar effect occurs,
when a self energy function according to (\ref{uvd}) is used,
i.e., when ghost poles would be present in a perturbative
propagator:  The dispersion integral always shifts complex (ghost)
poles to the unphysical Riemann sheet.

As a matter of fact it has been shown very long ago in
quantum field theory, that this is the only scenario compatible with
causality: The retarded as well as the advanced
propagator must be free of poles in the complex energy plane on
both sides of the real energy axis \cite{M60}.
%%%%%%%%%%%%%%%%%%%%%%%%%%%%%%%%%%%%%%%%%%%%%%%%%%%%%%%%%%%%%%%%%%%%%%%%%
\subsection{Approximate spectral functions}
The toy model spectral function of eqn. (\ref{toy}) cannot be used
in a relativistic calculation, since it propagates
only states of positive energy. More useful is a symmetric
ansatz, which for simplicity we first discuss for the bosonic case:
\begin{eqnarray}\nonumber
{\cal A}_B(E,\bbox{ k}) &  =  &
\frac{1}{2\omega_k}\,\left({\cal A}_0(E,\bbox{ k}) -
{\cal A}_0(-E,\bbox{ k})\right) \\
\label{ab}
&=&\frac{1}{\pi}
\,\frac{2 E \gamma_k}{\left(E^2-\Omega_k^2\right)^2+
4 E^2 \gamma^2_k}
\;,\end{eqnarray}
with  $\Omega^2_k  =  \omega_k^2+\gamma_k^2$.
This  spectral function is
normalized according to the canonical field commutation relations, i.e.,
\begin{equation}\label{normb}
\int\limits_0^\infty\!dE\,E\,{\cal A}_B(E,\bbox{ k}) =\frac{1}{2}
\;.\end{equation}
Note, that in this ansatz the dependence of $\gamma_k$ on the
momentum of the relativistic boson is completely arbitrary,
and one may also introduce a more general relationship between
$\omega_k$ and momentum than in the free case.

It is a matter of a few pages to obtain the full propagator from the
dispersion integral over this spectral function as
\begin{eqnarray}\nonumber
D^R(k_0,\bbox{ k}) &=& \frac{1}{2\pi\omega_k}
\left( \frac{\gamma_k}{(k_0-\omega_k)^2+\gamma_k^2}
      -\frac{\gamma_k}{(k_0+\omega_k)^2+\gamma_k^2} \right)
 \,\log(-k_0-{\mathrm i}\epsilon)\\
\nonumber
&+&\frac{1}{4\pi{\mathrm i}\omega_k}\left(
 \frac{\log(-\omega_k+{\mathrm i}\gamma_k)}{k_0-\omega_k+{\mathrm i}\gamma_k}
-\frac{\log(-\omega_k-{\mathrm i}\gamma_k)}{k_0-\omega_k-{\mathrm i}\gamma_k}
\right) \\
&-&\frac{1}{4\pi{\mathrm i}\omega_k}\left(
 \frac{\log(\omega_k+{\mathrm i}\gamma_k)}{k_0+\omega_k+{\mathrm i}\gamma_k}
-\frac{\log(\omega_k-{\mathrm i}\gamma_k)}{k_0+\omega_k-{\mathrm i}\gamma_k}
\right)
\;.\end{eqnarray}
This propagator is valid everywhere in the complex plane, and it is
by construction free of poles on the physical Riemann sheet. On the
real axis, it has the imaginary part $-\pi{\cal A}_B(k_0,\bbox{ k})$,
hence reproduces the ansatz for the spectral function.

The question is now, how one may relate such an approximate spectral function
to the general full propagator. Clearly it amounts to the
approximation of the denominator of a relativistic
retarded boson propagator along the real axis as
\begin{equation}\label{bosc}
k_\mu k^\mu - m^2 - \Pi^R(k) = (k_0-(\omega_k-{\mathrm i}\gamma_k))
                             \,(k_0+(\omega_k+{\mathrm i}\gamma_k))
\;,\end{equation}
where $\Pi$ is the boson self energy function,
and solving this equation for $\omega_k$, $\gamma_k$.
An example for such an approximation has been given in ref.
\cite{h94rep}, and there the momentum dependence of $\omega_k$
as well as of $\gamma_k$ was quite strong for the case of a pseudoscalar
coupling between bosons and fermions.

Let us remark at this point, that a {\em general\/} momentum dependence of
the $\gamma_k$-parameter does not guarantee the causality of the model,
i.e., it may violate the locality axiom as discussed above.
For a physically correct model, it is mandatory that the Fourier transform
of the spectral function vanishes for space-like coordinate arguments.
Indeed one may show, that for the ansatz eq. (\ref{ab}) this
is achieved when considering an $\omega_k^2$ which is quadratic in
$\bbox{k}$ and a constant $\gamma_k$.

This ansatz may be extended  to the fermionic case. The spectral function then
is slightly more complicated because of its Dirac matrix structure.
In eqn. (\ref{norm}) we already specified the normalization
for a fermion spectral function. The simplest ansatz compatible
with this normalization, as well as with known properties of fermion
systems, would be to multiply the spectral function
from eq. (\ref{ab}) by a factor $(E\gamma^0 + \omega_p)$.

However, as one may show, such an ansatz (used in an earlier version of the
present paper) violates the locality axiom, therefore is not useful for
our purpose. We therefore resort to a more complicated spectral function,
\begin{eqnarray} \nonumber
{\cal A}_F(E,\bbox{p}) &=& \frac{\gamma_p}{\pi} \vphantom{\int\limits_0^0}
\frac{\gamma^0\left(E^2 + \omega_p^2+ \gamma^2_p\right) +
      2 E \bbox{\gamma}\bbox{p} + 2 E M}{
  \left(E^2 - \omega_p^2 - \gamma^2_p\right)^2 + 4 E^2 \gamma_p^2}
\\ \nonumber
 &=&\frac{1}{4\pi{\mathrm i}\omega_p} \left(
\frac{\omega_p
  \gamma^0 + \bbox{p}\bbox{\gamma} + M}{E - \omega_p -
   {\mathrm i} \gamma_p}
-\frac{-\omega_p\gamma^0 + \bbox{p}\bbox{\gamma} + M}{
E+\omega_p
 - {\mathrm i} \gamma_p}\right.
\\
 &&\hphantom{\frac{1}{4\pi{\mathrm i}\omega_p}} \left.
-\frac{\omega_p\gamma^0 + \bbox{p}\bbox{\gamma} + M}{
E - \omega_p
 + {\mathrm i} \gamma_p}
+\frac{-\omega_p\gamma^0 + \bbox{p}\bbox{\gamma} + M}{
E+\omega_p
 + {\mathrm i} \gamma_p}
\right)\label{af}
\end{eqnarray}
where in principle $\gamma_p$ may be momentum dependent,
${\Omega_p}^2  =  {\omega_p}^2+{\gamma_p}^2 $
and $M$ is a constant which may be different from the physical
fermion mass in the vacuum state.

We now again turn to the question of consistency. As argued in the
previous subsection, the Fourier transform of the spectral function
has an important physical interpretation. We observe
that in the special case where $\gamma$ is independent of
momentum the 4-dimensional Fourier transform
of eq. (\ref{af}) is
\begin{equation}
C_F(x,0) =
  {\mathrm e}^{\displaystyle -\gamma
  \left|x_0\right|}\,
  \left({\mathrm i}\gamma^\mu\frac{\partial}{\partial x_\mu}+
  M\right)\,Z(x_0,\bbox{x})
\;,\end{equation}
with the function $Z$ as defined in eq. (\ref{c0def}).
Hence, for momentum independent $\gamma$ and $M$, the locality
axiom is fulfilled and interactions in the system are causal.
We therefore limit the use of the spectral function (\ref{af})
to this approximation, i.e.,
\begin{eqnarray} \nonumber
\gamma_p & \equiv\;\gamma&=\;\mbox{const.}\\
\nonumber
\omega_p^2 &\equiv\;\omega^2 & = \; \bbox{p}^2 + M^2\\
&\Rightarrow {\omega(\bbox{ p}+\bbox{ k})}^2 & = \;
  {\omega}^2 + 2 |\bbox{p}||\bbox{k}| \eta + k^2
\;,\end{eqnarray}
where $\eta$ is the cosine of the angle between the two momenta
$\bbox{p}$ and $\bbox{k}$.
%%%%%%%%%%%%%%%%%%%%%%%%%%%%%%%%%%%%%%%%%%%%%%%%%%%%%%%%%%%%%%%%%%%
\section{Fermion damping rate}
The next step is to use spectral functions, together
with the definition of the complete propagator by dispersion integral,
in a ''perturbative'' expansion at finite temperature.
We find, that ``perturbative'' is certainly not the correct label
for such a skeleton expansion, because in general due to the properties
of the dispersion integral the results will
{\em not\/} be expandable into power series in the coupling parameters
(see section II.C). Indeed up to correlation diagrams and
vertex corrections, a perturbative expansion in terms of
{\em full\/} propagators is exact already at the one-loop level
\cite[pp. 476]{IZ80}.

For simplicity, we begin the discussion with a simple scalar coupling
of a boson field having the spectral function ${\cal A}_B$ to
a fermion field having the spectral function ${\cal A}_F$.
In eqns. (\ref{vsconr}) and (\ref{psconr}), some
generalizations are investigated.

The ''one-loop'' diagram for the fermion self energy,
the Fock diagram, is given by the integral
\begin{eqnarray}\nonumber
&&\Sigma^{R}(p_0,\bbox{ p})  =\\
\label{retsi}
&&\;\;\;\;g^2\,\int\!\!\frac{d^3\bbox{ k}}{(2\pi)^3}\,
   \int\limits_{-\infty}^\infty\!\!dEdE^\prime\;
    {\cal A}_F(E,\bbox{ p}+\bbox{ k})\,{\cal A}_B(E^\prime,
\bbox{ k})\;
    \left(\frac{n_B(E^\prime)+n_F(E)}{
                p_0+E^\prime-E+{\mathrm i}\epsilon}\right)
\;.\end{eqnarray}
This function is split into real and imaginary part
as $\Sigma^R = \mbox{Re} \Sigma -{\mathrm i}\pi\Gamma$, and
for the latter we obtain
\begin{eqnarray}
\label{imsi}
&&\Gamma(p_0,\bbox{ p})  =\\
\nonumber
&&\;\;\;\;g^2\,\int\!\!\frac{d^3\bbox{ k}}{(2\pi)^3}\,
   \int\limits_{-\infty}^\infty\!\!dE\;
    {\cal A}_F(E+p_0,\bbox{ p}+\bbox{ k})\,{\cal A}_B(E,
\bbox{ k})\;
    \left(n_B(E)+n_F(E+p_0)\right)
\;.\end{eqnarray}
To fulfill the Kubo-Martin-Schwinger boundary condition (\ref{kmf}) for the
fermion propagator including the above self energy
function, it is {\em absolutely essential\/} that both
propagators in the above expression have the same equilibrium temperature.
In other words, one may {\em not\/} disregard the fermion distribution
function $n_F(E+p_0)$ in the above expression in any case, as
was done e.g. in ref. \cite{P93}. This fact will play an important role
in the next section as well as in sect. VI.B.

The corresponding ``one-loop'' result for the retarded scalar boson self energy
is, according to ref. \cite{h94rep}
\begin{eqnarray}\nonumber
\Pi^{R}(k_0,\bbox{ k}) &  = &
g^2\int\!\!\frac{d^3\bbox{ p}}{(2\pi)^3}\,
   \int\limits_{-\infty}^\infty\!\!dEdE^\prime \\ \label{retpi}
&&     \mbox{Tr}\left[
    {\cal A}_F(E,\bbox{ p}+\bbox{ k}){\cal A}_F(E^\prime,
\bbox{ p})
    \right]\,\left(\frac{n_F(E^\prime)-n_F(E)}{
                   k_0+E^\prime-E+{\mathrm i}\epsilon}\right)
\;.\end{eqnarray}
The imaginary part of $\Pi^R=\mbox{Re}\Pi-{\mathrm i}\pi\sigma$
is explicitly
\begin{eqnarray}\nonumber
\sigma(k_0,\bbox{ k}) &  = &
g^2\int\!\!\frac{d^3\bbox{ p}}{(2\pi)^3}\,
   \int\limits_{-\infty}^\infty\!\!dE \\ \label{sigpi}
&&     \mbox{Tr}\left[
    {\cal A}_F(E,\bbox{ p}+\bbox{ k}){\cal A}_F(E-k_0,
\bbox{ p})
    \right]\,\left(n_F(E-k_0)-n_F(E)\right)
\;.\end{eqnarray}
The above imaginary
part vanishes at $k_0=0$, more generally when the external energy
parameter equals the boson chemical potential. This
relation is {\em violated\/} if different propagators for the fermions are
used on the two legs of the diagram, e.g., when inserting different
``thermal masses'' for the two legs \cite{KLS91}.
This violation in turn leads to a violation of the
Kubo-Martin-Schwinger (KMS) boundary conditions for the
full boson propagator, i.e., it is {\em not\/} an equilibrium
Green function.

In principle, one could now introduce a self-consistent calculation scheme:
The expressions (\ref{retsi}) and (\ref{retpi})
are calculated with an approximate spectral function,
then used as input to eq. (\ref{sol3}) and a similar
equation for the full boson propagator in terms of its self energy.
This gives new spectral functions, which may be used
again to determine self energies. Such a scheme has been
used in refs. \cite{KM93,B84}, and based on these papers one might hope
that a few iterations are enough to determine the spectral function
numerically quite well.

However, as we have pointed out, there is no way to guarantee that
in such a scheme causality (in the representation of the locality
axiom) is preserved. Moreover, in an entirely numerical scheme
it would be impossible to point out the path to the solution of
the infrared problem. Two strategies exist which may be used to circumvent
this difficulty.

The first strategy is to enforce locality after each iterative step.
This can be done by folding the resultant
spectral functions with the Fourier transform of $\Theta(t^2-\bbox{x^2})$,
which makes this scheme numerically much more difficult than even the
iterative procedure used in \cite{KM93,B84}.

The second strategy is to use a simple causality-preserving
parametrization of the spectral functions. The parameters are then
determined self-consistently in a closed scheme. Apart from
the possibility to achieve analytical approximations, this strategy
also has the virtue that one can use it as an input to the first one.
Thus, for obvious reasons we will follow this second course.
%%%%%%%%%%%%%%%%%%%%%%%%%%%%%%%%%%%%%%%%%%%%%%%%%%%%%%%%%%%%%%%%%%%%%
\subsection{Approximations to the boson spectral function}
As was pointed out above, and is well known for many years,
the boson propagator may have isolated poles only on the real energy axis,
but not away from it on the physical Riemann sheet \cite{M60}.
Furthermore it may exhibit cuts along the real energy axis
due to self energy corrections with continuous nonzero imaginary part
like in eq. (\ref{sigpi}).

In absence of condensation phenomena these self energy corrections
have zero imaginary part when the boson energy parameter is equal
to the chemical potential.
Consequently, the product of a Bose-Einstein distribution function
(\ref{nbdf}) and a continuous part of the boson spectral function is always
{\em infrared finite}.

The only possible source for an infrared divergence therefore are poles
on the real axis, corresponding to particles with infinite lifetime.
We have argued before, that in principle such poles does not exist in
a finite temperature system \cite{NRT83}, and thus it is one on hand
clear that infrared divergences in finite temperature field theory
are not present in the full theory but a mere artifact of
perturbation theory.

On the other hand it is nevertheless instructive to study {\em how\/}
these divergences are removed within a proper calculational scheme.
We therefore chose the worst case for a boson spectral function, corresponding
to a kind of electromagnetic interaction without screening or damping:
\begin{equation}\label{abn}
{\cal A}_B^0 = \frac{1}{2 k} \,\left(\delta(E-k) - \delta(E+k)\right)
\;\end{equation}
with $k=|\bbox{ k}|$. It is obvious, that the product of this spectral
function with the corresponding (particle plus antiparticle)
Bose-Einstein distribution function is highly
singular for $E\rightarrow 0$.

However, inserting this spectral function into the calculation of $\Gamma$
gives at the lower boundary of the $k$-integration
\begin{equation}\label{inf1}
\Gamma(p_0,\bbox{p}) \sim g^2\,\int\limits_0\!\!\frac{dk}{(2\pi)^3}\,
\left(4 \pi\,2 T\,{\cal A}_F(p_0,\bbox{p}) + {\cal O}(k^2)\right)
\;,\end{equation}
i.e., the integral is {\em infrared finite\/} as long as the
fermion spectral function is finite. That it is also ultraviolet
finite is not obvious from the expansion made here, but follows from
eq. (\ref{imsi}).

We therefore conclude, that the use of a proper finite temperature
fermion propagator, which obeys the Narnhofer-Thirring theorem and therefore
has no isolated poles anywhere in the complex energy plane, is
completely sufficient to remove all infrared divergences from the
calculations. In particular, no screening or damping of the bosonic
interactions is needed to achieve this effect.

Introducing a modified (i.e., {\em less\/} singular) spectral function
for the bosons of course modifies the result for $\Gamma$
quantitatively, but only one direction for such a modification is
possible: The calculation of $\Gamma$ is dominated by the infrared
sector. Any less singular spectral function than (\ref{abn}) therefore
diminishes the value for $\Gamma$, because it leads to a
``smearing'' of the Bose pole with some distribution. Consequently,
we may consider the values obtained in the following an upper bound
on the full calculation.
%%%%%%%%%%%%%%%%%%%%%%%%%%%%%%%%%%%%%%%%%%%%%%%%%%%%%%%%%%%%%%%%%%%%%
\subsection{Approximations to the fermion spectral function}
We now set up a self-consistency scheme for the determination of
the fermion spectral function. Suppose, we really would approximate
the full spectral function according to eq. (\ref{sol3}) by the
simple parametrization of eq. (\ref{af}). This then would amount
to equate the denominators on the real energy axis as
\begin{eqnarray}\label{gamgam}\nonumber
&&\left(p_\mu - V_\mu(p) + {\mathrm i}\pi\Gamma^v_\mu(p)\right)^2 -
\left(M + S(p) -{\mathrm i}\pi\Gamma^s(p)\right)^2\\ & =&
\left(p_0 - \left(\omega_p - {\mathrm i}\gamma_p\right)\right)
\left(p_0 + \left(\omega_p + {\mathrm i}\gamma_p\right)\right)
\;.\end{eqnarray}
The functions of momentum appearing here are the components of the
retarded self energy function, split into real and imaginary part
according to
\begin{eqnarray}\label{sigs} \nonumber
\Sigma^R(p) &= & \mbox{Re}(\Sigma(p)) -{\mathrm i}\pi \Gamma(p)\\
\nonumber
\mbox{Re}(\Sigma(p)) &= & S(p) + V_\mu(p)\gamma^\mu\\
\Gamma(p) & = & \Gamma^s(p) + \Gamma^v_\mu(p)\gamma^\mu
\;.\end{eqnarray}
In the above expressions, we have also split the self energy functions
in a Lorentz scalar and a Lorentz vector piece, henceforth
simply abbreviated as scalar and vector part.

We have stated above, that in general this may give rise to a momentum
dependent $\gamma_p$ as well as to a non-quadratic dependence of
$\omega_p$ on $\bbox{p}$. Consequently, to check for the
locality axiom, i.e., the preservation of causality in the interactions
of the system, requires great numerical effort and one would lose
all the virtues that come with a simple parametrization
of ${\cal A}_F$. Let us therefore study first, how the choice of
spectral function (\ref{af}) affects the imaginary part of the self energy
(\ref{imsi}). Clearly it has only two independent components,
since its Lorentz scalar and vector part are
\begin{eqnarray}\nonumber
\Gamma^v_0(p)& =& \Gamma^{I}(p_0,\bbox{p},\gamma)\\
\nonumber
\Gamma^v_i(p)& =& \frac{\bbox{p}_i}{M}\,\Gamma^{II}(p_0,\bbox{p},\gamma)\\
\label{deco}
\Gamma^s(p)& =& \Gamma^{II}(p_0,\bbox{p},\gamma)
\;.\end{eqnarray}
In the following, we want to study a slow {\em massive\/} fermion on its
effective mass-shell, i.e., for $p_0=\omega_p=\sqrt{\bbox{p}^2 + M^2}$.
We therefore make the approximation that the real part of the self
energy function may be absorbed into the mass parameter $M$.

For the scalar coupling we consider here, the
self-consistency equation obtained from eq. (\ref{gamgam}) then reads
\begin{equation}\label{sconr}
\gamma_S=\pi\Gamma^{I}(\omega,\bbox{ p},\gamma_S)
+ \pi\Gamma^{II}(\omega,\bbox{ p},\gamma_S)\,
\left( 1- \frac{\bbox{p}^2}{M^2}\right)\,\frac{M}{\omega}
\;.\end{equation}
According to our assumption, the left side does
not depend on the momentum $\bbox{p}$. While we certainly cannot hope that
the momentum dependence on the right side cancels completely,
we may nevertheless expect that
it possesses an expansion of the type
\begin{equation}\label{ccc}
\gamma_S = \nu_0 + \nu_1 \frac{\bbox{p}^2}{M^2} +
\nu_2 \frac{\bbox{p}^4}{M^4} + \dots
\;.\end{equation}
Consequently, our approximations are reasonable if
the coefficients in this expansion are at least of the same
magnitude -- and they must be considered a failure when e.g.
$|\nu_0|\ll|\nu_1|$. To check this condition therefore will provide
an a-posteriori test of our assumptions.

It is quite simple to generalize this scheme to the exchange
of a massless vector boson, if the boson propagator is
taken in Feynman gauge \cite[pp.329]{IZ80}: The inclusion of
Dirac matrices $\gamma^\mu$, $\gamma_\mu$ at the vertices of the
Fock diagram simply multiplies $\Gamma^v$ by a factor -2 and
$\Gamma^s$ by a factor 4. The resulting
{\em vector boson} self consistency relation is
\begin{equation}\label{vsconr}
\gamma_V=-2\pi\Gamma^{I}(\omega,\bbox{ p},\gamma_V)
+ 4\pi\Gamma^{II}(\omega,\bbox{ p},\gamma_V)\,
\left( 1+ \frac{\bbox{p}^2}{2M^2}\right)\,\frac{M}{\omega}
\;.\end{equation}
It is worthwhile to note, that the ansatz of a momentum independent
imaginary part of the fermion self energy ensures the gauge
invariance of this self-consistent damping rate in the limit
$\bbox{p}\rightarrow 0$. Gauge invariance even of the first
momentum dependent correction term requires to introduce
a vertex correction in eq. (\ref{imsi}) to satisfy the correct
Ward identity \cite{h95curc}.

Physically interesting is furthermore the result of massless
{\em pseudoscalar bosons\/}, where one has
\begin{equation}\label{psconr}
\gamma_P=\pi\Gamma^{I}(\omega,\bbox{ p},\gamma_V)
-\pi\Gamma^{II}(\omega,\bbox{ p},\gamma_P)\,
\left( 1+ \frac{\bbox{p}^2}{M^2}\right)\,\frac{M}{\omega}
\;.\end{equation}
%%%%%%%%%%%%%%%%%%%%%%%%%%%%%%%%%%%%%%%%%%%%%%%%%%%%%%%%%%%%%%%%%%%%%
\subsection{Angular integrations}
The integration over the angle between the two momenta
$\bbox{k}$ and $\bbox{p}$ may be done analytically.
Using the above spectral functions, we obtain as the
$\gamma^0$-proportional imaginary part of the self energy
\begin{eqnarray}\nonumber
&&\Gamma^{I}(\omega,\bbox{ p},\gamma)  =\\
\nonumber
&&\;\;\;\;\frac{g^2}{4\pi^3}\,
\int\limits_0^\infty\!\!dk\left\{ \vphantom{\int}
k\gamma\,
\left( \gamma^0 \left( (k+\omega)^2-  k \omega + \gamma^2/2\right)
       \,I_1(k+\omega)
\right.\right.\\
\nonumber
&&\;\;\;\;\;\;\;\;\;\;\;\;\;\;\;\;\;\;\;\;
\left.
+ \gamma^0\,p k \,I_2(k+\omega) \right)
    \,\left(n_B(k)+n_F(k+p_0)\right)\\
\nonumber
&&\;\;\;\;\phantom{\frac{g^2}{4\pi^3}\,\int\limits_0^\infty\!\!dk}
-k\gamma\,
\left( \gamma^0 \left( (k-\omega)^2+ k \omega + \gamma^2/2\right)
       \,I_1(k-\omega)
\right.\\
\label{gamap1}
&&\;\;\;\;\;\;\;\;\;\;\;\;\;\;\;\;\;\;\;\;
\left.
+ \gamma^0\,p k \,I_2(k-\omega) \,\left(n_B(-k)+n_F(-k+p_0)\right)
\vphantom{\int}\right\}
\;.\end{eqnarray}
$I_1$ and $I_2$ are integrals over the angle between
the momenta $\bbox{p}$ and $\bbox{k}$, and we have abbreviated
$k=|\bbox{k}|$, $p=|\bbox{p}|$ (see appendix A for technical details).

A similar expression can be found for the second piece of the
self energy function,
\begin{eqnarray}\nonumber
&&\Gamma^{II}(\omega,\bbox{ p},\gamma)  =\\
\nonumber
&&\;\;\;\;M\frac{g^2}{4\pi^3}\,
\int\limits_0^\infty\!\!dk\left\{ \vphantom{\int}
k\gamma\, (k+\omega) \,I_1(k+\omega) \,\left(n_B(k)+n_F(k+p_0)\right)\right.\\
\label{gamap2}
&&\left.\;\;\;\;\phantom{\frac{g^2}{4\pi^3}\,\int\limits_0^\infty\!\!dk}
+k\gamma\, (k-\omega) \,I_1(k-\omega) \,\left(n_B(-k)+n_F(-k+p_0)\right)
\vphantom{\int}\right\}
\;.\end{eqnarray}
The infrared problem addressed above finds its representation in the
fact, that in the limit $\gamma\rightarrow 0$,
$I_1$ and $I_2$ behave like a {\em negative} power $\le -2$ of $k$.
This becomes obvious when expanding the integrand around the value $k=0$.

To check the conjecture of infrared finiteness made after eq. (\ref{inf1}), we
perform this expansion in the above expressions and
obtain for the quantity in curly brackets of eqn. (\ref{gamap1}),
at the point $p_0=\omega$ that we are specially interested in:
\begin{equation}\label{inf2}
\lim_{k\rightarrow 0} \left\{\vphantom{\int}\dots\right\}_{p_0=\omega}
=\gamma^0\frac{T}{\gamma}\,\frac{\omega^2+\gamma^2/2}{
\omega^2 + \gamma^2/4}
+{\cal O}(k^2)
\;.\end{equation}
Consequently, the infrared divergence is removed
by starting with a small nonzero $\gamma$.
%%%%%%%%%%%%%%%%%%%%%%%%%%%%%%%%%%%%%%%%%%%%%%%%
\section{Vacuum state}\label{GamVac}
The crucial difference between the present paper and hard thermal loop
calculations \cite{LS,P93,BNN93,PPS93,BPS94,BR94,W94,BK94,T94}
is, that we do not find it necessary to introduce momentum
cutoff scales: The quantity we are calculating is
infrared and ultraviolet finite, and therefore no justification
exists to neglect the ``vacuum'' contributions. These arise,
because the Bose-Einstein and Fermi-Dirac distribution functions
are nontrivial also for {\em zero\/} temperature, $n^0_B(E)=-\Theta(-E)$ and
$n^0_F(E)=\Theta(-E)$.

One may argue, that a ``particle'' in the vacuum generally is a well-defined
state, i.e., the NRT theorem \cite{NRT83} does not apply and therefore
the damping rate should be zero. This is of course correct, but we are
{\em not\/} using the ordinary vacuum propagators for the fermion field.
Instead, we must calculate the contributions with a nonzero
fermion damping rate as input, and only later may be able to
find a self-consistent solution of zero damping rate.

Using the above step functions in the
expression for the imaginary part of the self energy
leads to the simpler form
\begin{eqnarray}
&&\Gamma^{I}{\mbox{\tiny vac}}(\omega,\bbox{ p},\gamma)  = \\
\nonumber
&&\;\;\;\;\frac{g^2}{4\pi^3}\,
\int\limits_0^\infty\!\!dk \,
  k\,\gamma\,\left(
      \left( (k-\omega)^2+ k \omega + \gamma^2/2\right)
       \,I_1(k-\omega)+ p k  \,I_2(k-\omega)\right) \,\Theta(\omega-k)
\label{gamv1}
\;.\end{eqnarray}
This part of the self energy function has an asymptotic behavior according
to Weinberg's theorem \cite{W60}: The above integral is ultraviolet finite,
but the corresponding real part is divergent and has to be renormalized.
It is connected to the above
imaginary part only through a {\em subtracted\/} dispersion relation.
This asymptotic behavior of the self energy function
for large values of $p_0$ implies, that the ghost problem
discussed in sect. II.C becomes relevant.

In the light of our introductory remarks to this section
it is instructive to study the integral in case of zero $\gamma$
and momentum, where it becomes
\begin{eqnarray}\nonumber
\Gamma_{\mbox{\tiny vac}}^I(p_0,0,0) & =&
\frac{g^2}{4\pi^3}\,
\frac{\pi}{8}\,p_0\,\frac{(p_0^2+2 M p_0- M^2)^2
 (p_0^2-M^2)}{p_0^4 (p_0^2 + M^2)}\,
\Theta\left(\left|p_0\right|-M\right)\\
\label{gamv0}
&\rightarrow& \frac{g^2}{4\pi^3}\,
\frac{\pi}{8}\,p_0\;\;\mbox{if}\;\;
p_0\rightarrow\infty
\;.\end{eqnarray}
This imaginary part vanishes at $p_0=M$, more generally at
$p_0=\omega$ where we want to calculate
the spectral width. We may put this into the simple form:
Zero input $\gamma$ gives zero output $\Gamma_{\mbox{\tiny vac}}$
for ''on-shell'' particles.

However, this does no longer hold once we calculate the integral in
(\ref{gamv1}) with a nonzero positive $\gamma$ as input: Then,
we obtain a nonzero imaginary part also at the point
$p_0=\omega$.  In figure 1, we have plotted
$\Gamma^I_{\mbox{\tiny vac}}(p_0,0,\gamma)$ for two different values
of $\gamma$: At $p_0=M$, the imaginary part of the
Fock self energy function is zero only in case $\gamma=0$,
otherwise it is positive.

One may therefore complete the statement above: Nonzero input
$\gamma$ gives nonzero output $\Gamma_{\mbox{\tiny vac}}$
even at $p_0=\omega$.

We now exploit the virtue of a simply parametrized spectral
function for fermions, by expanding the continuous function
$\Gamma_{\mbox{\tiny vac}}$ around its value at $\gamma=0$.
As we have seen above, this value is zero.
However, it turns out that the imaginary part of the self
energy function with $p_0=\omega$ is not an
analytical function of $\gamma$: It has a cut in the complex
$\gamma$-plane, starting at $\gamma=0$. Thus, a Taylor expansion
in $\gamma$ is {\em not} possible, one has to perform an
asymptotic expansion, explicitly taking into account the leading
nonanalyticity.

Various techniques exist for such an expansion, for the purpose of
the present paper we isolated the leading terms by first
substituting $k\rightarrow x \gamma$ and then expanding the integrand
in powers of $\gamma$. It turns out, that for finite momentum $p$
the scale for the expansion is $\omega^2-p^2$ -- which implies,
that this asymptotic expansion is reliable only up to
momenta $p\approx M$.

Within this limitation, our result for the vector imaginary part
of the Fock self energy function is
\begin{eqnarray}\nonumber
\Gamma^{I}_{\mbox{\tiny vac}}(\omega,\bbox{p},\gamma) & = -
\displaystyle\frac{g^2}{4\pi^3}&\left\{
\frac{\gamma}{4}\,\frac{\omega^2}{\omega^2-p^2}\;
\left(\log\left(\frac{\gamma^2}{\omega^2-p^2}\right)
     -1+\frac{p}{\omega}\log\left(\frac{\omega+p}{\omega-p}\right)
\right)\right.\\
\nonumber
&&-
\frac{\gamma^2\,\pi}{8}\,
\frac{\omega (\omega^2+3 p^2)}{(\omega^2-p^2)^2}\\
\nonumber
&&+
\frac{\gamma^3}{24} \frac{1}{(\omega^2-p^2)^3}\,\left(
\vphantom{\int}
\frac{1}{6}\left(15 \omega^4 + 122 \omega^2 p^2 - 9 p^4\right)
 \right.\\
\nonumber
&&\;\;\;\;\;
\left.-
\left(9 \omega^4 + 20 \omega^2 p^2 + 3 p^4\right)
\,\log\left(\frac{\gamma^2}{\omega^2-p^2}\right)\right.\\
\label{gamva}
&&\;\;\;\;\;
\left.\left.-\frac{\omega}{p}\,
\left(2 \omega^4 + 18 \omega^2 p^2 + 12 p^4\right)
\,\log\left(\frac{\omega+p}{\omega-p}\right)
\right)\right\}+ {\cal O}(\gamma^4)
\;,\end{eqnarray}
where we have used $p=|\bbox{p}|$ and $k=|\bbox{k}|$.
In figure \ref{fig2}, this expansion in $\gamma$ is compared to
the full numerical calculation of eq. (\ref{gamv1}):
With each additional order that is included,
the quality of the approximation grows.

An important aspect of this expansion is the occurrence of
the quadratic term: In a naive view of the integral involved,
it is not present because $I_1(k-p_0)$ as well as
$I_2(k-p_0)$ are odd functions of $\gamma$.
However, as pointed out above, the function we are expanding is not
analytical in $\gamma$
-- and this effect causes the appearance of the quadratic
contribution. To demonstrate this, figure 2 also contains a curve where
this quadratic term has been left out.
Obviously in this case the third order
contribution makes the approximation worse instead of improving it.

In view of the above results we conclude, that for not too large
values of $\gamma$ the asymptotic expansion to {\em second order}
is well under control. Henceforth, to keep the results simple,
we therefore restrict the discussion to this next-to-leading log order.

We now turn to the second piece of the imaginary part of the
fermion self energy. According to (\ref{gamap2}) one obtains
in the vacuum
\begin{equation}\label{gamv2}
\Gamma^{II}_{\mbox{\tiny vac}}(\omega,\bbox{ p},\gamma)  =
-M\,\frac{g^2}{4\pi^3}\,
\int\limits_0^\infty\!\!dk \,
k\,\gamma\, (k-\omega) \,I_1(k-\omega) \,\Theta(\omega-k)
\;.\end{equation}
Performing the asymptotic expansion as above then gives
\begin{eqnarray}\nonumber
\Gamma^{II}_{\mbox{\tiny vac}}(\omega,\bbox{p},\gamma) & = -
\displaystyle\frac{g^2}{4\pi^3}&\left\{
\frac{\gamma}{4}\,\frac{M\omega}{\omega^2-p^2}\;
\left(\log\left(\frac{\gamma^2}{\omega^2-p^2}\right)
     +\frac{\omega}{p}\log\left(\frac{\omega+p}{\omega-p}\right)
\right)\right.\\
\nonumber
&&-
\frac{\gamma^2\,\pi}{8}\,
\frac{M(3\omega^2+ p^2)}{(\omega^2-p^2)^2}\\
\nonumber
&&+
\frac{\gamma^3}{8}\frac{M}{\omega(\omega^2-p^2)^3}\,\left(
\vphantom{\int}
\frac{1}{6}\left(21 \omega^4 + 14 \omega^2 p^2 - 3 p^4\right)
 \right.\\
\nonumber
&&\;\;\;\;\;
\left.- 4 \omega^2 (\omega^2 + p^2)
\,\log\left(\frac{\gamma^2}{\omega^2-p^2}\right)\right.\\
\label{gamsa}
&&\;\;\;\;\;
\left.\left.-\frac{\omega}{p}\,
\left(\omega^4 + 6 \omega^2 p^2 + p^4\right)
\,\log\left(\frac{\omega+p}{\omega-p}\right)
\right)\right\}+ {\cal O}(\gamma^4)
\;.\end{eqnarray}
We notice again, that as a sign of the
non-analyticity in the parameter $\gamma$,
a quadratic as well as a $\gamma\log(\gamma)$ contribution appear.
%%%%%%%%%%%%%%%%%%%%%%%%%%%%%%%%%%%%%%%%%%%%%%%%%%%%%%%%%%%%%%%%%%%%%%
\section{Thermal state}\label{GamTem}
We finally proceed by considering a thermal state, i.e., we
calculate the full imaginary part of the self energy function
according to (\ref{imsi}), which we split into the ``vacuum part''
calculated above and a temperature dependent part according to
\begin{eqnarray}\label{gts}\nonumber
\Gamma(p_0,\bbox{ p},\gamma)& =&
 \Gamma_{\mbox{\tiny vac}}(p_0,\bbox{ p},\gamma)
   + \Gamma_T(p_0\bbox{ p},\gamma) \\
&=& \Gamma_{\mbox{\tiny vac}}(p_0,\bbox{ p},\gamma)
+\Gamma_T^I(p_0,\bbox{ p},\gamma)\gamma^0 +
 \Gamma_T^{II}(p_0,\bbox{ p},\gamma)
\,\left(1+\frac{\bbox{p}\bbox{\gamma}}{M}\right)
\;,\end{eqnarray}
similar to the decomposition for the vacuum part only in eq. (\ref{deco}).
For the moment it seems quite hopeless to obtain an {\em analytical\/}
approximation to the full expression (\ref{gamap1}) valid for
all temperatures. However, we know that for not too high temperatures
the Bose-Einstein and Fermi-Dirac distribution function may be
expanded around their zero-temperature value,
\begin{eqnarray}\label{disa}\nonumber
n_B(k) + n_F(p_0+k) &\approx&
  \exp(-k/T)\left(\frac{T}{k} + \frac{1}{2} + n_F(p_0)
  + \frac{k}{T}\left(\frac{1}{12} + n_F(p_0)^2\right)\right)\\
\nonumber
n_B(-k) + n_F(p_0-k) & \approx &
  - \exp(-k/T)\left(\frac{T}{k} + \frac{1}{2} + \frac{k}{T}\frac{1}{12}
  \right)\\
\nonumber
- \exp(-k/T)\mbox{e}^{p_0/T}&\cdot&\Theta(k-p_0)
\left( n_F(p_0) +\frac{k}{T}n_F(p_0)
(1-n_F(p_0))\right)\\
&+& \Theta(p_0-k)\left(-1 + n_F(p_0) +\frac{k}{T}n_F(p_0)
(1-n_F(p_0))\right)
\;.\end{eqnarray}
The term with the $-1$ in the last line constitutes
the ``vacuum'' part as obtained in the previous section.
The other terms are ordered
according to their dominance, i.e., the contribution
proportional to $T/k$ is the largest due to the strongly peaked
nature of the function $\exp(-k/T)$. The terms including the
fermionic distribution functions are negligible for $p_0\gg T$.

It was stated in eqns. (\ref{inf1}) and (\ref{inf2}), that the
introduction of a small nonzero $\gamma$ removes the
infrared divergence from the integral. Thus, similar to
the vacuum case, one obtains a
result for $\Gamma_T$ which cannot be expanded in a Taylor series
around the point $\gamma=0$.

We have described above how to obtain an asymptotic expansion
around this value. Here this is again achieved
by rescaling $k\rightarrow x \gamma$ in the integrand,
and then expanding everything but the exponential factor $\exp(-x \gamma/T)$
in powers of $\gamma$.

To improve the calculation, yet another trick is introduced:
First, the momentum integral is calculated using only the leading term
$T/k = 1/(\beta k)$ of the above expansion. The result is then improved
by acting on it with a differential operator
\begin{equation}\label{gamvt2}
{\Gamma_T^I}^\prime(\omega,\bbox{ p},\gamma)
= \Gamma_T^I(\omega,\bbox{ p},\gamma)
  -\left(\frac{1}{2}\frac{\partial}{\partial \beta}
        -\frac{1}{12 T} \frac{\partial^2}{\partial \beta^2}
  \right)\,\frac{\Gamma_T^I(\omega,\bbox{ p},\gamma)}{T}
\;.\end{equation}
For both pieces of the self energy function this is a quite laborious
task, for our calculation we heavily relied on symbolic computation
using Mathematica${}^{\mbox{\tiny tm}}$. For the first piece one obtains
\begin{eqnarray}\label{gamvt3}\nonumber
&&{\Gamma_T^I}^\prime(\omega,\bbox{ p},\gamma)=
  \displaystyle\frac{g^2}{4\pi^3}
\left\{ \frac{\pi T \omega}{4 p}\,\log\left(
     \frac{\omega+p}{\omega-p}\right) \right.\\
\nonumber
&&+\frac{\gamma}{4}\,\frac{\omega^2}{\omega^2-p^2}\,
    \left(\log\left(\frac{\gamma^2 \omega^2}{
    T^2(\omega^2-p^2)}\right)
%%\right. \\
%%\nonumber
%%&&\left.\hphantom{+\frac{\gamma}{4}\,\frac{\omega^2}{\omega^2-p^2}}
+ \frac{\omega}{p}\log\left(\frac{\omega+p}{\omega-p}\right) +
    2 C_\Gamma - \frac{17}{3}
    + \frac{20 T^2}{3 \omega^2}\right)\\
\nonumber
&&-\frac{\gamma^2 \pi}{24 T}\,
   \frac{\omega^4 + 18 \omega^2 T^2 + 18 T^2}{(\omega^2-p^2)^2}\\
\nonumber
&&+\frac{\gamma^3}{(\omega^2-p^2)^2}\left(
  \frac{(\omega-p)^5}{8p}\,\left(
   \log\left(\frac{\gamma\omega}{T (\omega+p)}\right)
   + C_\Gamma - \frac{19}{12}\right)\right.\\
\nonumber
&&\hphantom{\frac{\gamma^3}{(\omega^2-p^2)^2}}
   -\frac{(\omega+p)^5}{8 p}\,\left(
   \log\left(\frac{\gamma\omega}{T (\omega-p)}\right)
   + C_\Gamma - \frac{19}{12}\right)\\
&&\;\;\;\;\;\left.\left.
   \vphantom{\int}
   +\left(3\omega^2+p^2\right)
    \left(\frac{\omega^4}{216 T^2}
     -\frac{5}{9} T^2\right)\right)\right\}
   + {\cal O}(\gamma^4)
\;,\end{eqnarray}
where $C_\Gamma = 0.57721...$ is Euler's constant.
A comparison of this approximation to different orders in
$\gamma$ with the full numerical calculation of the
temperature dependent contribution is presented in
fig. \ref{fig4}.
As before, it turns out that the results are already quite
reliable in second order of $\gamma$, provided the temperature
is not too high.

In case of the vacuum parts of the self energy function we
were able to gather all terms of a given order in $\gamma$ in this
asymptotic expansion. For the present temperature dependent pieces
this is {\em not\/} possible, because of the above expansion:
It automatically counts $T$ and $\gamma$ to be of the same
order, hence is a simultaneous expansion in these parameters.
As a small reminder of this fact appears a term $\gamma T^2$
in the above expansion, and the ${\cal O}(\gamma^4)$ must
be replaced by a ${\cal O}(\gamma^4,\gamma T^3)$

We find, that in second order $\gamma$ the
asymptotic expansion is reliable up to $\gamma\approx T/2$, whereas
it is good up to the temperature in third order.
In anticipation of the following section, we emphasize that the
quality of the approximation grows tremendously for very small
temperatures $T\ll 0.1 M$

With similar quality we may also obtain the
second piece, asymptotically given by
\begin{eqnarray}\label{gamst3}\nonumber
&&{\Gamma_T^{II}}^\prime(\omega,\bbox{ p},\gamma)=
  \displaystyle\frac{g^2}{4\pi^3}
\left\{ \frac{\pi T M}{4 p}\,\log\left(
     \frac{\omega+p}{\omega-p}\right) \right.\\
\nonumber
&&+\frac{\gamma}{4}\,\frac{M \omega}{\omega^2-p^2}\,
    \left(\log\left(\frac{\gamma^2 \omega^2}{
    T^2(\omega^2-p^2)}\right)
%%\right.\\
%%\nonumber
%%&&\left.
%%\hphantom{\frac{\gamma}{4}\,\frac{M \omega}{\omega^2-p^2}\,}
+ \frac{\omega}{p}\log\left(\frac{\omega+p}{\omega-p}\right) +
    2 C_\Gamma - \frac{17}{3}\right)\\
\nonumber
&&-\frac{\gamma^2 \pi M \omega}{24 T}\,
   \frac{\omega^2 + 12 T^2}{(\omega^2-p^2)^2}\\
\nonumber
&&+\frac{\gamma^3 M}{(\omega^2-p^2)^3}\left(
   \frac{(\omega-p)^4}{8 p}\,\left(
   \log\left(\frac{\gamma\omega}{T (\omega+p)}\right)
   + C_\Gamma - \frac{19}{12}\right)\right.\\
\nonumber
&&\hphantom{\frac{\gamma^3 M}{(\omega^2-p^2)^3}}
-\frac{(\omega-p)^4}{8p}\,\left(
   \log\left(\frac{\gamma\omega}{T (\omega-p)}\right)
   + C_\Gamma - \frac{19}{12}\right)\\
&&\;\;\;\;\;\left.\left.+
 \frac{\omega^3}{216 T^2} \left(3 \omega^2 + p^2\right)
    -\frac{1}{16 \omega} \left(\omega^4 + 6 \omega^2 p^2 + p^4\right)
  \right)\right\}
+ {\cal O}(\gamma^4,\gamma T^3)
\;.\end{eqnarray}
To this expression one may apply the same considerations as before:
Although there is no piece $\propto \gamma T^2$, we may trust
this expansion in second order $\gamma$ only up to $\gamma\approx T/2$.

The claims we lay on the accuracy of our asymptotic expansions
are supported by fig. \ref{fig4}, where we compare them
with a completely independent numerical calculation
We have checked the accuracy of our results with various values for
momentum and temperature, and found it to persist up to momenta
$p\approx M$.

To be completely sure of our findings, we furthermore
will drop the corrections of orders $\gamma^3$: Unusual as it may
seem for quantum field theory, we therefore have an approximate method
with controlled accuracy in the next-to-leading-log order.
%%%%%%%%%%%%%%%%%%%%%%%%%%%%%%%%%%%%%%%%%%%%%%%%%%%%%%%%%%%%%%%%%%%
\section{Fermion damping rate II}\label{scfa}
Finally, we assemble the expressions for the self energy
function that we have obtained in the previous two sections.
Combining (\ref{gamva}) with (\ref{gamvt3}) and
(\ref{gamsa}) with (\ref{gamst3}),
we then observe that {\em all the terms of order\/}
$\gamma\,\log(\gamma)$ {\em cancel !\/}

This occurs for scalar and vector part of the self energy
function independently, hence hold for all types
of massless bosons that were considered in the equations
(\ref{sconr}) - (\ref{psconr}). We will comment on this in more detail
below.

As a consequence of this fact,
the self-consistent damping rate for the fermion moving slowly
through a hot medium, as defined in the equations (\ref{sconr})
-- (\ref{psconr}), is the solution of an {\em algebraic equation}
of the form
\begin{equation}\label{scon}
f_0(\omega,\bbox{ p},T)  - \left( f_1(\omega,\bbox{ p},T)\,
 +\frac{4 \pi}{g^2}\right)\, \gamma +
f_2(\omega,\bbox{ p},T)\,\gamma^2 = 0
\;.\end{equation}
This equation has always two solutions for $\gamma$,
but in general one of them
is negative or very large. Hence it is obvious how to choose the physical
solution of the above equation, and in the following we are referring
to this one only.

Our results do not exclude the possibility of corrections of
order $\gamma^3 \log(\gamma)$ to this equation.
The coupling constant appears only at one point,
the three functions $f_0$ - $f_2$ do not depend on it.
$f_0$ is proportional to the temperature $T$, $f_1$ contains
terms which are not analytical around zero temperature, i.e.,
terms of order $T\,\log(T)$.

The solution $\gamma$ of eqn. (\ref{scon}) therefore is
a power series in the coupling constant -- but still it includes
a non-perturbative effect in form of the non-analytical behavior in
the temperature, which is of order $g^4\,T\,\log(T)$.

One might argue, that these non-analytical terms are
not significant because in order $g^4$
other two-loop diagrams also give a contribution that we have neglected.
However, as may be expected from our results, their
non-analytical contribution involving $\log(T)$ is suppressed
with respect to their leading order contribution by a factor $\gamma$,
which may be translated into a factor $g^2 T$.
This implies, that our results to order $g^4$
yield the {\em dominant\/} non-analyticity and are important for
small temperatures (see figures).

Moreover, this chain of arguments also supports the conclusion that
terms of order $g^2 T p^2/M^2 \log(1/g)$ cannot appear: Terms logarithmic
in the coupling constant wear at least a coefficient $g^6$ or
$g^4 T^2/M^2$, even if vertex corrections are introduced in the
calculation of the self energy function.

Figs. \ref{fig6} and \ref{fig7} depict the self consistent
solution of eq. (\ref{scon})
for scalar bosons coupled to the fermions. For small
temperatures, the deviation from the linear temperature
dependence is very striking: Due to the $T\log(T)$ terms in the
self-energy functions, $\gamma_S(T)$ rises sharply at very
small temperature (see comment in the next section).

Before discussing this in detail, we turn to the results for
the vector and pseudoscalar boson exchange.
The self-consistent $\gamma_V$ due to massless vector boson exchange
in Feynman gauge is plotted in figs. \ref{fig8} and \ref{fig9}.
The curves basically employ the same features as for the scalar bosons:
A pronounced non-analyticity of the function $\gamma_V(T)$ at
small temperatures.

For the pseudoscalar boson our approach leads to a result which is
surprising at the first glance: The self-consistent solution for
$\gamma_P$ is identically zero at all temperatures. This follows
from the fact that the leading terms in the temperature dependence
of eqs. (\ref{gamvt3}) and (\ref{gamst3}) differ only by a factor
of $M/\omega$. Inspection of eq. (\ref{psconr}) then reveals that
these leading terms cancel, consequently $f_0\equiv 0$ in the
above equation (\ref{scon}) and therefore $\gamma_P\equiv 0$.

However, at a second glance this result is not surprising.
The pseudoscalar coupling is a p-wave coupling, hence vanishes
for zero momentum transfer. This is reflected in the fact that
the coefficients $f_1$ and $f_2$ start with power $p^2$ instead
of possessing a constant piece. We therefore expect a self-consistent
$\gamma_P$ which also starts with this power $p^2$ -- clearly in contradiction
to our ansatz spectral function.

Naturally it is possible to repeat the calculation with a corresponding
ansatz for $\gamma$ -- but this would overstress the goals of the present
paper. We therefore use this pseudoscalar result merely as a consistency
argument in favor of our approximations.

Also another feature of our results for $\gamma_S$ and $\gamma_V$
may be seen in the figures.
Clearly, the momentum dependence of our self-consistent solution is small
up to quite high temperatures. In view of eq. (\ref{ccc}) this means
$|\nu_0| \raisebox{-1ex}{$\stackrel{>}{\sim}$} |\nu_1|$
for scalar and vector boson
exchange. We therefore conclude, that the ansatz of
a momentum independent $\gamma_S$ and $\gamma_V$ is
very well justified.
%%%%%%%%%%%%%%%%%%%%%%%%%%%%%%%%%%%%%%%%%%%%%%%%%%%%%%%%%%%%%%%%%%%%
\subsection{Analytical approximations for the damping rate}
It is not necessary to repeat the analytical expressions for
the self energy pieces in order to give the three functions
$f_0$ -- $f_2$ in closed form. However, it is
instructive to extract the dominant terms in the
calculation of $\gamma$.

To this end we perform an expansion of the solution of
(\ref{scon}) in powers of the coupling constant.
Only even powers occur, hence the expansion parameter is
\begin{equation}
\alpha = \frac{g^2}{4 \pi}
\;.\end{equation}
The contributions of order $T$ and of order $T\,\log(T)$ then
are, to second order in $\alpha$
\begin{eqnarray}\nonumber
\gamma_S & \approx &  \alpha T \,\left(
1 - \frac{2}{3} \frac{p^2}{M^2} + \frac{8}{15}\frac{p^4}{M^4}\right)\\
\nonumber
& -& \alpha^2 \frac{T}{\pi}
  \,\left(\log\left[\frac{T}{M}
    \left(1-\frac{11}{12}\frac{p^2}{M^2}+
    \frac{1061}{1440}\frac{p^4}{M^4}\right)\right]
  +\frac{25}{12} - C_\Gamma\right)
 \,\left(1 - \frac{2}{3} \frac{p^2}{M^2} +
 \frac{8}{15}\frac{p^4}{M^4}\right)
\\ \label{lead}
&+& {\cal O}\left(
   \alpha\frac{p^6}{M^6},
   \alpha^2\frac{T^2}{M^2},
   \alpha^3\right)
\;,\end{eqnarray}
where $p=|\bbox{p}|$.
Naturally this last expansion is only good in the weak coupling
limit. Performing the same expansion for the vector boson
exchange yields
\begin{eqnarray}\nonumber
\gamma_V & \approx &   \alpha T \,\left(
1 - \frac{2}{3} \frac{p^2}{M^2} + \frac{8}{15}\frac{p^4}{M^4}\right)\\
\nonumber
& -& \alpha^2 \frac{T}{\pi}
  \,\left(\log\left[\frac{T}{M}
    \left(1-\frac{1}{3}\frac{p^2}{M^2}+
    \frac{31}{180}\frac{p^4}{M^4}\right)\right]
   +\frac{13}{3} - C_\Gamma\right)
 \,\left(1 - \frac{2}{3} \frac{p^2}{M^2} +
 \frac{8}{15}\frac{p^4}{M^4}\right)
\\ \label{leadv}
&+& {\cal O}\left(
   \alpha\frac{p^6}{M^6},
   \alpha^2\frac{T^2}{M^2},
   \alpha^3\right)
\;.\end{eqnarray}
It is a crucial aspect of these results, that they are
analytical functions of the coupling constant, i.e., in contrast
to other authors we do not find contributions of order
$\log(g^2)$ \cite{P93,PPS93,BK94}.

During the discussion of the self-consistency criterion
(\ref{vsconr}) we have already stated, that the vector boson
result is gauge independent
in the limit $\bbox{p}\rightarrow 0$. A vertex correction is needed
to ensure gauge invariance for the momentum dependent parts (which in principle
also require a completely different calculation scheme).
However, this will at most give a regular modification of order
$\alpha^2 p^2/M^2$, logarithmic terms may appear only in even higher
order $\alpha$.

The contribution of order $T\,\log(T)$, apart from being non-analytical
around $T=0$, has the effect of destroying the linear relationship
between $\gamma$ and $T$: Neither for high temperatures nor for
low temperatures can one approximate the self-consistent $\gamma$
by a linear function. This functional piece constitutes the
leading non-analyticity observed numerically in figures
(\ref{fig6}) -- (\ref{fig9}).
%%%%%%%%%%%%%%%%%%%%%%%%%%%%%%%%%%%%%%%%%%%%%%%%%%%%%%%%%%%%%%%%%%%%%
\subsection{Where does the log go ?}
A striking feature of our result is, that for
a slow massive fermion there are no contributions to the
damping rate of order $g^2 T \log\left(1/g\right)$.
This seems to contradict some existing papers on the subject of
the fermion damping rate, eg. refs. \cite{BPS94,BK94}, whereas
others clearly denounce the idea of a logarithmic correction
\cite{PPS93,T94}. We therefore have to set our result in relation to
such modifications.

Since our study was to a large extent motivated by ref.
\cite{BK94}, we take the liberty to quote their eq. (15),
with $e$ replaced by $g$:
\begin{eqnarray}\label{bk94}
\gamma &\simeq& \frac{g^2}{2 \pi^2}\,T\,\int\limits_0^{g T}\!\frac{dk}{k}\,
\arctan\left(\frac{k}{\gamma}\right)\\
&\simeq&\frac{g^2}{4 \pi}\,T\,\int\limits_\gamma^{g T}\!\frac{dk}{k}
\label{bk94ff}
\;\simeq\;\frac{g^2}{4\pi}\,T\,\log\left(\frac{g T}{\gamma}\right)\\
\label{bk94f}
&\simeq&\frac{g^2}{4 \pi}\,T\,\log\left(\frac{1}{g}\right)
\;\;\;\mbox{according to ref. \cite{BK94}.}
\end{eqnarray}
Before we discuss the difference of this approach to ours, let us
first point out that also in our scheme appears a similar expression:
In the calculation of $\Gamma^I_T$ as well as in $\Gamma^{II}_T$ appear
terms of the type
\begin{equation}
\frac{g^2}{4\pi^2}\,\int\!\!dk\,\frac{T}{k}\,
 \arctan\left(\frac{2 k \omega \pm 2 p k - \gamma^2}{2\gamma (\omega\pm
  k)}\right)\,{\mathrm e}^{-k/T}
\;.\end{equation}
They are due to the combination of angular
integral factors (\ref{i1i}) with
the leading term in the expansion of the Bose-Einstein distribution function
(\ref{disa}). Restricting the parameters contained here to the
approximations $ \bbox{p}=0, k\ll \omega, \gamma^2 \ll 2 k \omega$,
one indeed obtains the same integrand as in eq (\ref{bk94}).

Two questions arise now: First of all it is important to ask,
whether the ``standard'' result from eq. (\ref{bk94f}) is
correct in itself, i.e., whether the final expression really is
an approximation to the first expression. To answer this
in detail, we note that one may indeed evaluate the integral in
eq. (\ref{bk94}) in closed form. It is obtained as the
Lerch transcendent, a
generalization of the polylogarithm function and not
expressible as a finite series of elementary functions.

However, its numerical evaluation is straightforward, and we find that
with reasonable accuracy it is indeed approximated by
the expression in (\ref{bk94ff}) -- but {\em not\/} by the final expression
(\ref{bk94f}). Instead, it follows from (\ref{bk94}), that
\begin{equation}\label{bk94imp}
\gamma \simeq \frac{g^2}{4\pi}\, T\,\log\left(\frac{1}{g}\right)\,
  \left(1 - \displaystyle \log\left[\log\left( \frac{1}{g}\right)
 \frac{1}{4 \pi}\right]\left\{\log\left(
  \displaystyle\frac{e}{g}\right)\right\}^{-1} \right)
\;,\end{equation}
where $e=2.71828\dots$.
As can be seen from fig. 8, this makes a tremendous difference: Even
for coupling parameters as small as in QED, the step from
(\ref{bk94ff}) to (\ref{bk94f}) is not justified, and
the difference between the two expressions is {\em not\/} of higher order.

The second question, equally
relevant for the present paper,
is the consistency of the first line in eq. (\ref{bk94}).
To this end we notice that in our approach there is no restriction to
the momentum $k$: The $k$-integration is automatically cut off at
a scale $k\approx T$ due to the exponential function, see
the expansion in eq. (\ref{disa}).

Even more, as we have found the zero temperature pieces
(involving $\Theta$-functions) of the distribution functions are
crucial in replacing the non-analyticity in the coupling constant
by a non-analyticity in the temperature. The answer to the second
question therefore is, that the first line in eq. (\ref{bk94})
(and similar expressions) may be consistent in the framework of
hard thermal loop calculations, but is not a good approximation to the
fully self-consistent damping rate problem.

We may summarize the difference of the two approaches by
symbolically writing down
two different approximation schemes for the distribution functions
and the corresponding result for the imaginary part of the self
energy function:
\begin{eqnarray}\nonumber
\mbox{hard thermal loop resummation:}\hfill&&\\
\nonumber
\left. { \array{lll}
n_B(k) + n_F(\omega+k) & \approx & \frac{T}{k}\,\Theta(g T- k) \\
n_B(-k) + n_F(\omega-k) & \approx & -\frac{T}{k}\,\Theta(g T+ k)
\endarray } \right\} & \Rightarrow & \frac{g^2}{4 \pi}\, T\,
 \log\left(\frac{g T}{\gamma}\right)\\
\nonumber
\mbox{this work:}\hfill\\
\label{htlf}
\left. { \array{lll}
n_B(k) + n_F(\omega+k) & \approx &
\left(\frac{T}{k} +\frac{1}{2}\right)\,
\exp(-k/T) \\
n_B(-k) + n_F(\omega-k) & \approx &-\left(\frac{T}{k} +\frac{1}{2}\right)\,
\exp(-k/T)\\
&& -\Theta(\omega-k)
\endarray } \right\} & \Rightarrow & \frac{g^2}{4 \pi}\, \left(T\,
 +\gamma\log\left(\frac{\omega}{T}\right)\right)
\;.\end{eqnarray}
Naturally, in our approximation scheme one may no longer use
$k\ll\omega$ but has to perform the full calculation as presented in
the foregoing sections. We therefore conclude at this point, that the absence
of terms of order $g^2 T \log(1/g)$ in our analytical approximations
cannot be used as a reliable argument against our work.
%%%%%%%%%%%%%%%%%%%%%%%%%%%%%%%%%%%%%%%%%%%%%%%%%%%%%%%%%%%%%%%%%%%
\section{Conclusion}
With the present paper we have reached several goals. First of all,
we have presented a straightforward application of the method of
generalized free fields. This automatically accounts for the
undeniable mathematical fact that at nonzero temperature a perturbative
expansion in terms of quasi-particle states with infinite lifetime
is ill-defined \cite{NRT83,BS75,L88}.

As we have shown, this method
is also practical in the sense that it removes any unphysical
infrared divergence from a perturbative expansion. We may put this into the
following form: In reality a fermion in a heat bath is always subject to
some Brownian motion, hence will never be in the same state
long enough to emit a very soft boson.

The central ingredient of the method are (more or less) continuous
spectral functions. These functions are not a priori fixed in a theory
involving generalized free fields \cite{L88}, one may determine
them from experiment or use physically reasonable self-consistency
schemes. One such scheme we have presented here: In assigning
a constant spectral width to fermions, we go one step beyond
the introduction of temperature dependent (``thermal'') masses.

This approximation is reasonable for systems which are dominated
by the low-momentum sector of the interactions, i.e., those which
involve massless gauge bosons. Naturally, to determine
transport properties of plasmas in true non-equilibrium states, all the details
of the spectral function have to be known and the approximation
of a constant damping rate will break down \cite{h94rep}.

For the massless bosons we used a ``free'' spectral function,
without even a constant spectral width parameter. Although, in the spirit
of our previous remarks, one may expect this to lead to inconsistencies
in higher orders of perturbation theory we have made this choice for two
particular reasons. First of all our calculation
clearly shows, that and exactly how this worst-case infrared divergent
interaction (which includes the Coulomb singularity)
is regularized by the use of a nontrivial {\em fermion\/} spectral function.
Secondly, any more realistic spectral functions for the bosons will only
weaken the infrared behaviour of this spectral function, therefore
leads to a {\em smaller\/} self-consistent damping rate.

Numerical calculations show, that using such damped bosonic interactions in the
self-consistent calculation is equivalent to inserting
the sum of the fermionic $\gamma$ and the bosonic $\gamma_B$
on the right side of eq. (\ref{sconr}), i.e., in the calculated
imaginary parts of the self energy function. Such
a substitution does {\em not\/} affect the lowest order
thermal contribution as obtained in sect. V. of this work.
Thus it must be concluded, that the use of an effective boson
propagator does not change the leading order
result for the fermionic spectral width.

For our approximate fermion spectral function we have achieved the
summation of all nested Fock diagrams to arbitrary order. Apart from
vertex corrections, such a summation constitutes the solution of
the full one-body problem. As we have pointed out, such vertex corrections
would lead to momentum dependent corrections to the damping
rate - and therefore would not affect the momentum independent parts
of our results.

In the weak coupling limit these momentum independent parts are
analytical functions of the coupling constant,
which sets them apart from some published
results obtained with the method of hard thermal loops. We therefore
made a careful investigation of the reason for logarithmic corrections
within this method, and found them to be due to a certain approximation
made to the Bose-Einstein and Fermi-Dirac distribution function.
A comparison to the approximation used in the present
work is summarized in eq. (\ref{htlf}).

We find that logarithmic (i.e., $\log(1/g)$)
corrections may appear only in higher orders of the self-consistent
calculation, i.e., when including the self energy function
to order $\gamma^3\log(\gamma)$. As we have demonstrated by our
use of a ``free'' boson propagator, they are certainly not due
to the absence of screening on boson interaction lines.

Another crucial difference of the calculations presented here
to those established in the literature is the use of
entirely causal propagators. Our spectral functions
obey the locality axiom of quantum field theory (sect. II.B and D),
whereas numerical studies show that this is {\em not\/} the case
for those generally used in ``hot gauge theory'' \cite{h95comm}.
One may argue, that this violation is of minor importance for
the problem of the damping rate -- but in our view the
interaction of a fermion with its own radiation field should be
{\em strictly\/} causal.

Instead of a non-analyticity in the coupling constant,
our results exhibit a non-analyticity in the temperature
parameter arising from contributions of order $T\log(T)$.
This finding is in accordance with the symmetry of space and time:
In the presence of temperature the Poincar\'e symmetry is broken
to SO(3)$\times$T${}_4$, and the symmetry restoration with
temperature $T\rightarrow 0$ is expected to be singular.

We must stress the fact, that the calculation of the real part
of the self energy function by a dispersion integral was suppressed in
the present paper. According to the arguments we presented,
the dispersion integral may still lead to non-analytical contributions
in the coupling constant. However, these we expect to
be of order $\exp(-1/g^2)$ due to the removal of the Landau ghost poles
\cite{h92fock}, and therefore they are varying slowly around the zero
point of the coupling constant.

A final conclusion to be drawn from our paper is, that it is certainly
dangerous to present non-analytical results, like e.g. of order
$g^4 T \log(T)$, without elaborate cross-checks. We have solved this
problem by performing numerical calculations along with the
asymptotic expansion of loop integrals, hence each figure
presented here contains results obtained by two independent
methods. Furthermore, every expansion made was carried out to one
order higher than was finally used.
\subsection*{Acknowledgment}
The results presented here have been critically checked and discussed
for more than one year. We owe our particular thanks to
B.Friman, J.Knoll, E.Quack, L.Razumov and D.Voskresensky.
%%%%%%%%%%%%%%%%%%%%%%%%%%%%%%%%%%%%%%%%%%%%%%%%%%%%%%%%%%%%%%%%%%%
\appendix
\section{Angular integration}
In the following we give the explicit form of the angular integrals
over the approximate fermion spectral function, which occur in the
calculation of (\ref{gamap1}) and (\ref{gamap2}).
$\eta$ in this appendix resembles the cosine of the angle between the two
momenta $\bbox{k}$ (internal) and $\bbox{p}$ (external),
and we will henceforth write $k=|\bbox{k}|$, $p=|\bbox{p}|$.
\begin{eqnarray}\nonumber
I_1(k\pm p_0) & = & \int\limits_{-1}^1\!d\eta\,
  \frac{1}{(c_1- 2 p k \eta)^2 + c_2^2}\\
\label{i1i}
  & = & \frac{-1}{4pk(k \pm p_0)\gamma}\left.
 \arctan\left(\frac{(k \pm p_0)^2-t^2-\gamma^2}{
               2(k \pm p_0)\gamma}\right)
\right|_{\omega_{p-k}}^{\omega_{p+k}}
\end{eqnarray}
and
\begin{eqnarray}
I_2(k\pm p_0) & = & \int\limits_{-1}^1\!d\eta\,
  \frac{\sqrt{c_3+ 2 pk \eta}}{(c_1- 2 p k \eta)^2 + c_2^2}\\
\nonumber
&=&\frac{c_1}{2 k p}\,I_1(k\pm p_0)
+\frac{1}{8 p^2 k^2}\left.
 \log\left(\frac{\left(t^2+\gamma^2-(k\pm p_0)^2\right)^2 +
                4\gamma^2(k\pm p_0)^2 }{\gamma^2}\right)
\right|_{\omega_{p-k}}^{\omega_{p+k}}
\end{eqnarray}
with
\begin{eqnarray}\nonumber
c_1 & =& \pm2 p_0 k + p_0^2 - {\omega}^2 - {\gamma}^2\\
\nonumber
c_2 & =& 2 (k\pm p_0)\gamma\\
c_3 & = & {\omega}^2 + k^2
\end{eqnarray}
and boundaries of the integration defined as
\begin{equation}
\omega_{p\pm k}^2 = \omega^2 \pm 2 p k + k^2
\;.\end{equation}
An important aspect of the above analytical results
is to select the proper Riemann sheet for the arctan-functions:
$I_1$ and $I_2$ are continuous functions in each variable.
%%%%%%%%%%%%%%%%%%%%%%%%%%%%%%%%%%%%%%%%%%%%%%%%%%%%%%%%%%%%%%%%%%

\clearpage
\begin{figure}
\caption{Imaginary part $\Gamma^I_{\mbox{\tiny vac}}$
of the self energy function.}
Temperature $T=0$, momentum $p=0$, factor $g^2/(4 \pi^3)$ set to one.\\
Different constant values of $\gamma$:
Thin line: $\gamma=0$, eqn. (\ref{gamv0});
thick line: $\gamma=0.2M$
\hrule
\label{fig1}
\end{figure}
%%%%%%%%%%%%%%%%%%%%%%%%%%%%%%%%%%%%%%%%%%%%%%%%%%%%%%%%%%%%%%%%
\begin{figure}
\caption{Imaginary part $\Gamma_{\mbox{\tiny vac}}$:
Numerical calculation vs. analytical approximation.}
Top panel: vector piece $\Gamma^I_{\mbox{\tiny vac}}$;
bottom panel: scalar piece $\Gamma^{II}_{\mbox{\tiny vac}}$\\[1mm]
Temperature $T=0$, energy $p_0=\omega$,
momentum $p=0.5 M$, factor $g^2/4\pi^3$ set to one.\\[1mm]
Continuous thin line: Full numerical calculation.\\[1mm]
Dash-Dotted thick line: Expansion to order $\gamma$.\\
Dashed thick line: Expansion to order $\gamma^2$.\\
Continuous thick line: Expansion to order $\gamma^3$.\\
Dotted thin line: Expansion to order $\gamma^3$, but
{\em without\/} the term of order $\gamma^2$.
\label{fig2}\label{fig3}
\hrule
\end{figure}
%%%%%%%%%%%%%%%%%%%%%%%%%%%%%%%%%%%%%%%%%%%%%%%%%%%%%%%%%%%%%%
\begin{figure}
\caption{Imaginary part $\Gamma^\prime_T$:
Numerical calculation vs. analytical approximation.}
Top panel: vector piece ${\Gamma^I_T}^\prime$;
bottom panel: scalar piece ${\Gamma^{II}_T}^\prime$.\\[1mm]
Energy $p_0=\omega$, momentum $p=0.5 M$, factor $g^2/(4\pi^3)$ set to one.\\
Temperatures are $T=0.05 M$ (bottom lines) and $T=0.1 M$ (top lines).\\
Thin continuous lines: Full numerical calculation.\\
Dash-dotted thick lines: Asymptotic expansion to order $\gamma$\\
Dashed thick lines: Asymptotic expansion to order $\gamma^2$\\
Thick continuous lines: Expansion to order $\gamma^3$.
\label{fig4}\label{fig5}
\hrule
\end{figure}
%%%%%%%%%%%%%%%%%%%%%%%%%%%%%%%%%%%%%%%%%%%%%%%%%%%%%%%%%%%%%%%
\begin{figure}
\caption{Self-consistent $\gamma_S/M$ as function of $T/M$}
Strong coupling $\alpha$ = 1, compared to $\gamma=T$ (thin line). \\
Two different momenta:  $p=0$ (thick continuous) $p=0.5 M$
(thick dashed)\\[1mm]
\hrule
\label{fig6}
\end{figure}
%%%%%%%%%%%%%%%%%%%%%%%%%%%%%%%%%%%%%%%%%%%%%%%%%%%%%%%%%%%%%%%%%
\begin{figure}
\caption{$(\gamma_S -  \alpha T)/M$ as function of $T/M$}
Top panel: strong coupling $\alpha$ = 1;
bottom panel: weak coupling $\alpha$ = 0.1\\[1mm]
Continuous lines: $p=0$; dashed lines: $p=0.5 M$
\label{fig7}
\hrule
\end{figure}
%%%%%%%%%%%%%%%%%%%%%%%%%%%%%%%%%%%%%%%%%%%%%%%%%%%%%%%
\begin{figure}
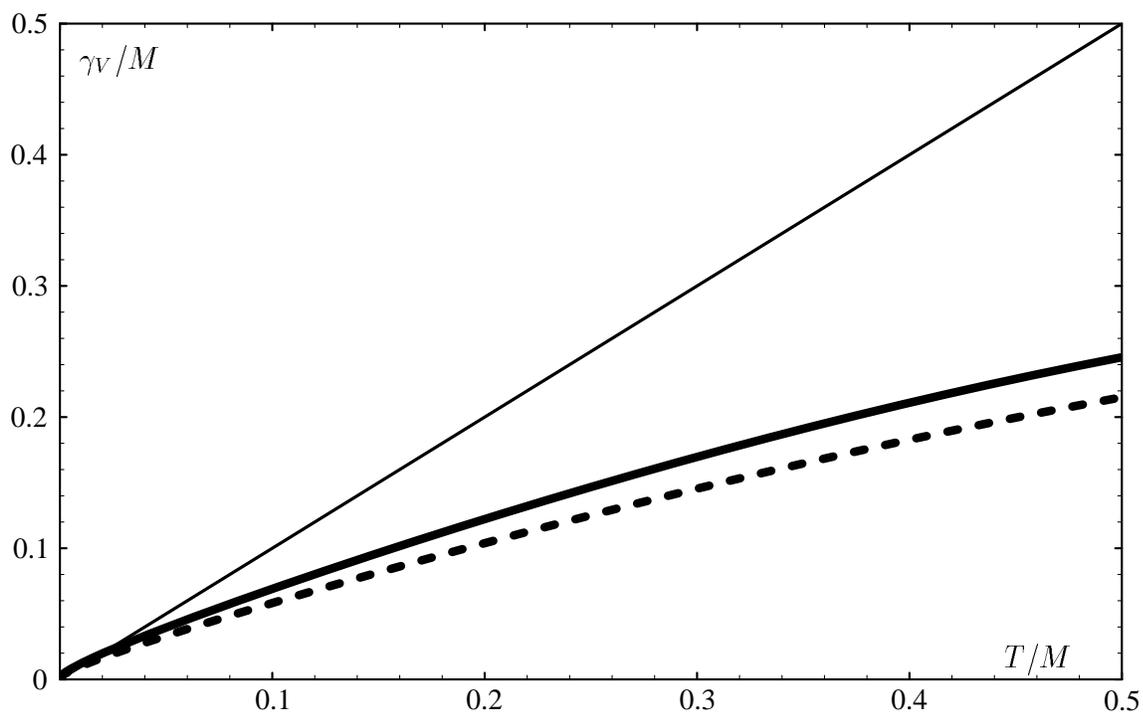

\caption{Self-consistent $\gamma_V/M$ as function of $T/M$}
Strong coupling $\alpha$ = 1, compared to $\gamma=T$ (thin line). \\
Two different momenta:  $p=0$ (thick continuous) $p=0.5 M$
(thick dashed)
\hrule
\label{fig8}
\end{figure}
%%%%%%%%%%%%%%%%%%%%%%%%%%%%%%%%%%%%%%%%%%%%%%%%%%%%%%%%%%%%%
\begin{figure}
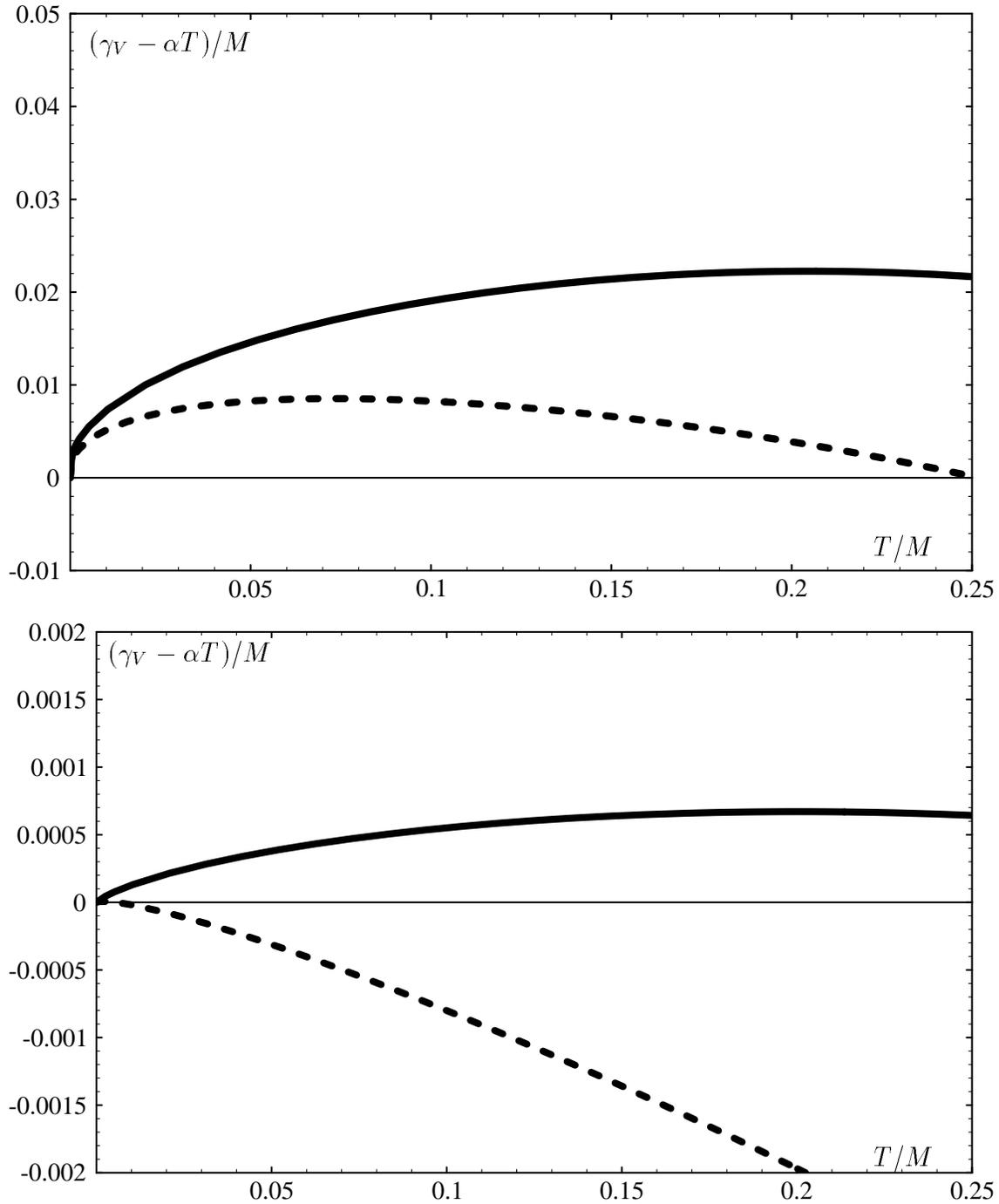

\caption{$(\gamma_V -  \alpha T)/M$ as function of $T/M$}
Top panel: strong coupling $\alpha$ = 1;
bottom panel: weak coupling $\alpha$ = 0.1\\[1mm]
Continuous lines: $p=0$; dashed lines: $p=0.5 M$
\label{fig9}
\hrule
\end{figure}
%%%%%%%%%%%%%%%%%%%%%%%%%%%%%%%%%%%%%%%%%%%%%%%%%%%%%%%%%%%%%%%%
\begin{figure}
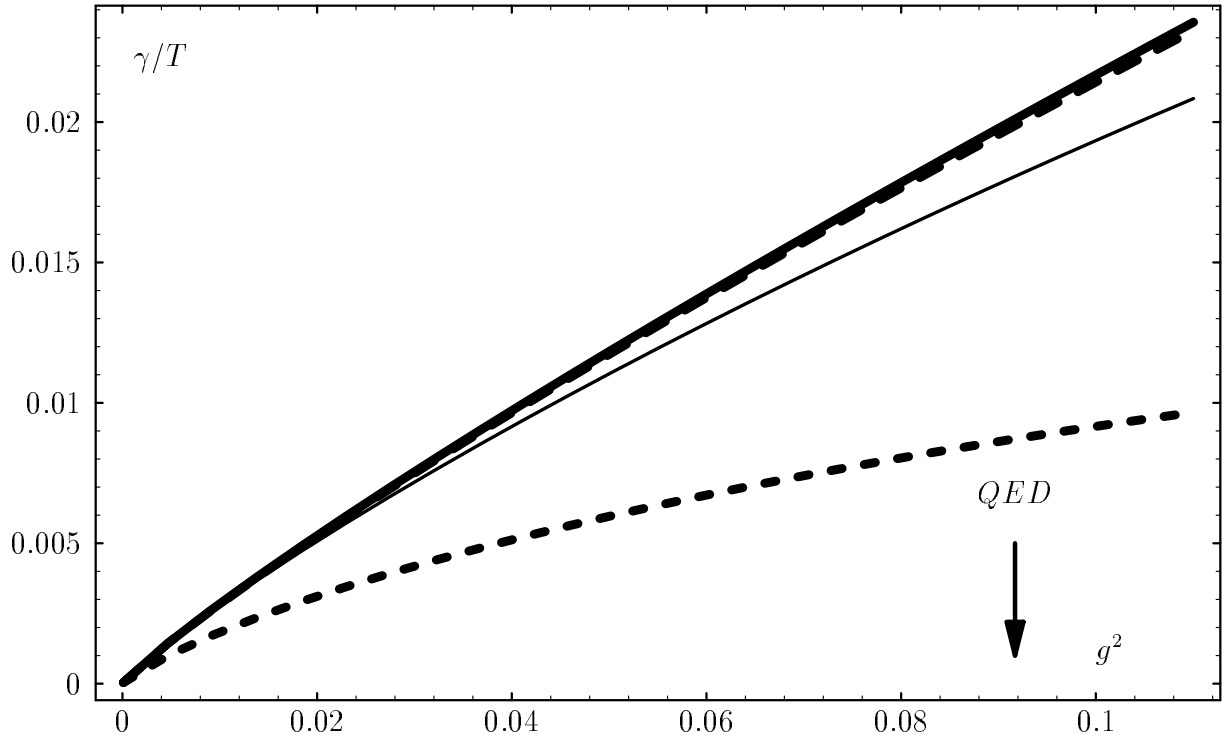

\caption{Comparison of analytical approximations to eq. (\ref{bk94}).}
Continuous thick line: numerical calculation of (\ref{bk94}).\\
Upper dashed line: approximation according to (\ref{bk94ff}).\\
Lower dashed line: approximation according to (\ref{bk94f}).\\
Continuous thin line: approximation according to (\ref{bk94imp}).\\[1mm]
\label{fig10}
\hrule
\end{figure}

\clearpage
\begin{thebibliography}{99}
\bibitem{LS}{
    V.V.Lebedev and A.V.Smilga, Ann.Phys. {\bf 202} (1990) 229,\\
    Phys.Lett. {\bf B253} (1991) 231,  Physica {\bf A 181} (1992) 231}
\bibitem{P93}{
    R.D.Pisarski, Phys.Rev. {\bf D47} (1993) 5589}
\bibitem{BNN93}{
    R.Baier, H.Nakkagawa and A.Ni\'egawa, Can.J.Phys. {\bf 71} (1993) 205}
\bibitem{PPS93}{
    S.Peign\'e, E.Pilon and D.Schiff,
    Z.Physik {\bf C60} (1993) 455}
\bibitem{BPS94}{
  R.Baier, S.Peign\'e and D.Schiff,
    Z.Phys. {\bf C62} (1994) 337}
\bibitem{BR94}{
    M.Le Bellac and P.Reynaud,\\
    {\em in\/} Proceedings of the 3rd Workshop on Thermal
    Field Theories and their Applications,\\
    eds. F.C.Khanna, R.Kobes, G.Kunstatter and H.Umezawa\\
    (World Scientific, Singapore 1994) p.440}
\bibitem{W94}{
    H.A.Weldon,
    {\em in\/} Proc. 3rd Workshop on TFT, ibid. p. 450\\
    and Phys.Rev. {\bf D49} (1994) 1579}
\bibitem{BK94}{
    R.Baier and R.Kobes,Phys.Rev. {\bf D 50} (1994) 5944}
\bibitem{T94}{
    M.H.Thoma, Z.Phys. {\bf C66} (1995) 491}
\bibitem{hq95gam}{
    P.A.Henning and E.Quack, Phys.Rev.Lett {\bf 75} (1995) 2811}
\bibitem{h95neq}{
    P.A.Henning, hep-ph 9510315; to appear
    {\em in\/} Proceedings of the 4th Workshop on Thermal
    Field Theories and their Applications,\\
    eds. Y.X.Gui and F.C.Khanna
    (World Scientific, Singapore, t.b.p. 1996)}
\bibitem{BP90}{
    E.Braaten and R.D.Pisarski,
    Nucl.Phys. {\bf B337} (1990) 569}
\bibitem{NRT83}{
    H.Narnhofer, M.Requardt and W.Thirring,\\
    Commun.Math.Phys. {\bf 92} (1983) 247}
\bibitem{BS75}{
    H.J.Borchers, R.N.Sen,
    Commun.Math.Phys.{\bf 21} (1975) 101}
\bibitem{L88}{
    N.P.Landsman,\\
    Phys.Rev.Lett. {\bf 60} (1988) 1990 and
    Ann.Phys. {\bf 186 } (1988) 141}
\bibitem{L65}{
   A.L.Licht, Ann.Phys. {\bf 34} (1965) 161}
\bibitem{W67}{
   A.S.Wightman, {\em in:\/}
   { Carg\`ese Lectures in Theoretical
   Physics:\\  High Energy Electromagnetic Interactions and
   Field Theory,}\\  ed. M. L\'evy
   (Gordon \& Breach, New York 1967)}
\bibitem{LW87}{
    N.P.Landsman and Ch.G.van Weert,
    Phys.Rep. {\bf 145} (1987) 141}
\bibitem{SKF}{
    J.Schwinger,
    J.Math.Phys. {\bf 2} (1961) 407;\\
    L.V.Keldysh, Zh.Exsp.Teor.Fiz. {\bf 47} (1964) 1515 and
    JETP {\bf 20} (1965) 1018}
\bibitem{Ubook}{
    H.Umezawa,\\
    { Advanced Field Theory: Micro, Macro and Thermal Physics}\\
    (American Institute of Physics, 1993)}
\bibitem{h94rep}{
    P.A.Henning,
    Phys.Reports {\bf 253} (1995) 235}
\bibitem{hu92}{
    P.A.Henning and H.Umezawa,\\
    Nucl.Phys. {\bf B417} (1994) 463; Phys.Lett. {\bf B 355} (1995) 241}
\bibitem{KMS}{
    R. Kubo,
    J.Phys.Soc. Japan {\bf 12} (1957) 570;\\
    C.Martin and J.Schwinger,
    Phys.Rev. {\bf 115} (1959) 1342}
\bibitem{KS85c}{
    R.L.Kobes and G.W.Semenoff,\\
    Nucl.Phys. {\bf B260} (1985) 714;
    Nucl.Phys. {\bf B272} (1986) 329}
\bibitem{KV95}{
    J.Knoll and D.Voskresensky,
    Phys.Lett. {\bf B351} (1995) 43 and\\
    GSI-Preprint 95-63, subm. to Ann.Phys.}
\bibitem{h95comm}{
    P.A.Henning, E. Poliatchenko and T.Schilling,\\
    {\em Approximate spectral functions in thermal field theory}\\
    GSI Preprint 95-70 (1995); hep-ph 9510322}
\bibitem{GJ81}{
    J.Glimm and A.Jaffe, Quantum Physics\\
    (Springer, New York 1981)}
\bibitem{W60}{
    S.Weinberg, Phys.Rev. {\bf 118} (1960) 848}
\bibitem{IZ80}{
    C.Itzykson and J.B.Zuber,\\
    { Quantum Field Theory }
    (McGraw-Hill, New York 1980)}
\bibitem{h92fock}{
    P.A.Henning,
    Nucl.Phys. {\bf A546} (1992) 653}
\bibitem{M60}{
    S.Mandelstam, Nuovo Cim. {\bf 15} (1960) 658}
\bibitem{KLS91}{
    J.Kapusta, P.Lichard and D.Seibert,
    Phys.Rev. {\bf D44} (1991) 2774}
\bibitem{KM93}{
    C.M.Korpa and R.Malfliet,
    Phys.Lett. {\bf B315} (1993) 209}
\bibitem{B84}{
    A.F.Bielajew and B.D.Serot,
    Ann.Phys. {\bf 156} (1984) 215}
\bibitem{h95curc}{
    P.A.Henning and M.Blasone, GSI Preprint 95-36 (1995),
    hep-ph 9507273}
\end{thebibliography}
\end{document}